\documentclass{aa}  

\usepackage{pdfpages}
\usepackage{graphicx}
\usepackage{xcolor}
\usepackage{makecell} 
\usepackage{txfonts}
\usepackage{lipsum}
\usepackage{multirow}
\usepackage{todonotes}
\usepackage{lscape}

\renewcommand{\sun}{$_{\odot}$}
\newcommand{\teff}{T$_{\rm eff}$\,}
\newcommand{\met}{[M/H]\,}
\newcommand{\logg}{$\log{g}$\,}
\newcommand{\gssp}{\textsc{GSSP}\,}
\newcommand{\kms}{$\rm km \ s^{-1}$}
\newcommand{\vsin}{$v\sin{i}$\,}
\newcommand{\vmic}{$v_{\rm mic}$\,}
\newcommand{\vmac}{$v_{\rm mac}$\,}

\begin{document} 

   \title{Wide sdB binaries. I. Orbital and atmospheric parameters}

   \author{
         Francisco Molina\inst{1,2}
        \and 
        Joris Vos\inst{3}
        \and
        Alexey Bobrick\inst{4,5}
        \and
        Maja Vu\v{c}kovi\'{c}\inst{6}
    }

   \institute{
        Institut f\"{u}r Physik und Astronomie, Universit\"{a}t Potsdam, Karl-Liebknecht-Str. 24/25, 14476, Golm, Germany
        \and 
        Dpto. Química "Prof. José Carlos Vílchez Martín", Facultad de CC Experimentales, Universidad de Huelva, 21007 Huelva, Spain
        \and
        Astronomical Institute of the Czech Academy of Sciences, CZ-25165, Ond\v{r}ejov, Czech Republic
        \and
        School of Physics and Astronomy, Monash University, Clayton, Victoria 3800, Australia
        \and
        ARC Centre of Excellence for Gravitational Wave Discovery -- OzGrav, Australia
        \and 
        Instituto de F\'{\i}sica y Astronom\'{\i}a, Universidad de Valpara\'{\i}so, Gran Breta\~{n}a 1111, Playa Ancha, Valpara\'{\i}so 2360102, Chile
        }

   \date{}

  \abstract
   {Long-period binary systems containing a B-type hot subdwarf (sdB) and a main-sequence (MS) companion are thought to originate from binary interactions involving stable mass transfer from the red giant, the progenitor of the sdB, to the MS companion. However, despite the recent progress in modelling their population, some of their observed properties are not entirely understood. Because determining their orbits requires extended campaigns of high-resolution spectroscopic observations, only a limited number of long-period sdB binaries have been studied with completely determined orbital parameters. }
   {We aim to expand the current sample of long-period sdB binaries with fully determined orbital parameters through the analysis of high-resolution spectroscopic data. In addition, the atmospheric parameters of the cool companions are analysed. Increasing the number of well-characterized systems will provide valuable insights into their formation channels and main characteristics.}
   {A sample of 32 wide binary systems containing sdB stars was selected for the analysis of the radial velocity (RV) curves of both companions. The dataset consisted of high-resolution spectra obtained with the HERMES and UVES spectrographs. The orbital parameters were derived by simultaneously fitting Keplerian orbits to the RVs of the sdB and its companion. The atmospheric parameters of the cool companions were determined using the GSSP code, which analyses the master spectra of the systems with a grid of LTE atmospheric models. 
   An additional sample of wide sdB binaries was built up by cross-matching the Gaia NSS catalogue with catalogues of sdB candidates and spectroscopically confirmed systems reported by \citet{Culpan2022A&A}. The outcomes from both samples were compared with existing theoretical models to assess their consistency with current formation and evolutionary scenarios.}
   {The complete orbital solution of the 32 wide sdB binaries was obtained. Their Galactic orbits and kinematics parameters were derived. Likewise, we provide the set of atmospheric parameters of the cool companions.}
   {}


   \maketitle
%

\section{Introduction}
\label{sec:Introduction}

Binary interactions and mass transfer processes lead to the formation of numerous types of post-interaction binaries, including hot subdwarf B stars, which populate the extreme horizontal branch (EHB). These sdB stars are classically described as core-helium-burning objects with a thin hydrogen envelope (approximately 0.01 M\sun) and a mass near the canonical core-helium-flash value of 0.47 M\sun \ \citep{Saffer1994ApJ,Brassard2001ApJ, Heber2009ARA&A,Heber2016PASP}, although the actual mass range is likely broader \citep{ArancibiaRojas24}. They are characterized by high effective temperatures (20,000 to 40,000 K) and surface gravities typically ranging from 5.0 to 6.0 dex. In the classical scenario, an sdB star forms from the progenitor that lost most of its hydrogen envelope near the tip of the red giant branch and ignited the core through a helium flash \citep[RGB, e.g.,][]{Heber2016PASP}.

It is now widely accepted that all sdB stars form exclusively through binary interactions \citep{Heber2016PASP,Pelisoli2020A&A}, as initially suggested by the early work of \citet{Mengel1975BAAS}. The three main formation channels proposed to date are: the common-envelope (CE) ejection channel \citep{Paczynski1976IAUS,Han2002MNRAS}, the stable Roche-lobe overflow (RLOF) channel \citep{Han2000MNRAS,Han2002MNRAS}, and the formation of a single sdB star through the merger of two helium white dwarfs \citep[WD, ][]{Webbink1984ApJ}.

The RLOF channel can form long-period sdB binaries with orbital periods exceeding 500 days, typically containing main-sequence stars as cool companions \citep{Vos2020A&A}. Observations of such systems have been first reported in the past decade \citep{Ostensen2012ASPC, Deca2012MNRAS, Vos2012A&A}. In particular, some observed orbital periods surpassed the predictions of existing binary population synthesis (BPS) models \citep{Han2002MNRAS, Han2003MNRAS}, highlighting the need for an updated theoretical framework. Subsequently, \citet{Chen2013MNRAS} proposed an atmospheric RLOF model to explain periods of up to 1600 days. 

The most recent modelling showed good agreement with the observed formation rates, orbital periods, mass ratios, and metallicities of the wide composite sdB population \citep{Vos2020A&A}.
However, some relevant features of these systems are not entirely explained by the models to date. A notable case is the eccentricity at long orbital periods. The work of \citet{Vos2015A&A} proposed a theoretical framework to address this issue, based on the combination of two eccentricity pumping mechanisms: phase-dependent RLOF and a circumbinary (CB) disc. Despite this, the model falls short in reproducing the observed trend in longer orbital periods and is unable to predict the eccentricity of the systems.

Increasing the number of accurately determined orbital parameters for long-period sdB binaries will help constrain our understanding of their formation, properties, and evolution.

The article is organised as follows. Sect.\,\ref{sec:Sample and observations} describes the main characteristics of the wide sdB binary samples used in this work and summarises the observations' information, based on either ground-based data or Gaia. Sect.\,\ref{sec:Orbital} outlines the determination of the orbital solutions and parameters for both samples. The analysis of the atmospheric parameters of the cool companions in wide sdB systems is presented in Sect.\,\ref{sec:AP_MS}. The Sect.\,\ref{sec:Properties} provides the information on the Galactic orbits and kinematics derived parameters of the systems. Subsect.\,\ref{sub:models} establishes the main parameter relations and compares the results with the current theoretical models. A discussion has been developed on the possible causes of agreement or disagreement between observations and models.
Finally, Sect.\,\ref{sec:conclusions} summarises the main findings and conclusions of this work.


\section{Sample and observations}
\label{sec:Sample and observations}

Two different samples of wide sdB binaries are considered in this article. The first one consists of known sdBs that are part of our long-term observing program.
Their collection of high-resolution spectra spans more than fifteen years and involves observations from the {\sc Mercator} telescope at the Roque de los Muchachos Observatory \citep[La Palma, Canary Islands, Spain;][]{Gorlova2013EAS,Vos2012A&A,Vos2013A&A,Vos2017A&A}. The program was extended to cover southern targets using the Very Large Telescope at Paranal Observatory \citep[{\sc VLT}; Atacama Desert, Chile;][]{Vos2018MNRAS,Vos2019MNRAS}.

The second sample consists of wide sdB binaries observed by Gaia \citep{Gaia2023A&A674_A1}, for which orbital solutions have been derived from either Radial Velocity Spectrometer (RVS) data or astrometric measurements. The subsequent sections outline the target selection criteria and the observational data for both samples.


\subsection{Ground-based sample}
\label{sub:Ground}

This sample consists of known wide sdB binaries whose spectral data were obtained with HERMES (High Efficiency and Resolution Mercator Echelle Spectrograph) at the Mercator telescope and with UVES (Ultraviolet and Visual Echelle Spectrograph) at the VLT. The target candidates were selected from a multitude of low- and medium-resolution spectroscopic and photometric surveys, including but not limited to \citet{Green1986ApJS,Downes1986ApJS,Kilkenny1988SAAOC, Stark2003AJ, Rhee2006BA,Wade2006BA}. The first target selection was performed using the original hot subdwarf database compiled by \citet{Ostensen2006BA}, which contains more than 2300 stars. The target list was updated when new bright sdB/Os were found in later surveys. The original surveys and their more recent follow-up observing campaigns have been consolidated in the hot subdwarf catalogues of \citet{Geier2017A&A, Geier2020A&A} and \citet{Culpan2022A&A}. 

The final targets were selected based on their magnitude. The northern hemisphere targets with a limiting magnitude of 11.5 in the V band were observed with HERMES, while in the southern hemisphere, targets up to V=14 were included, since a larger telescope was used. Binarity was determined based on radial velocity (RV) variations in the data/spectra. 
HERMES targets were followed for at least  2 years before determining binarity, usually involving more than 10 observations.
The UVES targets were followed over the course of two observing seasons (6-12 months), discarding from the sample those sources not showing significant variation.

HERMES is a high-resolution spectrograph mounted on the 1.2m Mercator Telescope at the Roque de los Muchachos Observatory in La Palma, Spain. 
It has a spectral resolution of R = 85000, and covers a spectral range from 377 to 900 nm \citep{Raskin2011A&A}. 
It is designed for precise spectroscopic studies of stars managed under conditions of high wavelength stability. 

There are 10 targets observed with HERMES between 2009 and 2017: 8 sdBs and 2 long-period binaries containing an evolved O-type hot subwarf (sdO). A detailed discussion of the HERMES observations can be found in \citet{Vos2017A&A} for the sdB binaries and \citet{Molina2022A&A} for the sdO ones, along with technical information regarding the spectrograph capabilities. The HERMES sample was completed, and no new data have been obtained.

UVES is a high-resolution spectrograph installed on the 8.2m  VLT at the Paranal Observatory in Chile. UVES is a two-arm cross-dispersed echelle spectrograph, which was used in standard dichroic-2 437+760 mode, covering a wavelength range of 373–499 nm in the BLUE arm and 565–946 nm in the RED arm. The resolutions achieved for each arm, using a slit width of 1 arcsecond, are R = 41,000 and 42,000, respectively.
The spectrograph is designed to study the chemical composition, kinematics, and physical conditions of astronomical objects.  
The complete set of observations by HERMES and UVES is summarized in Table \ref{21-Obs} of the appendix.

\begin{figure*}
  \centering
  \includegraphics[width=8.5cm]{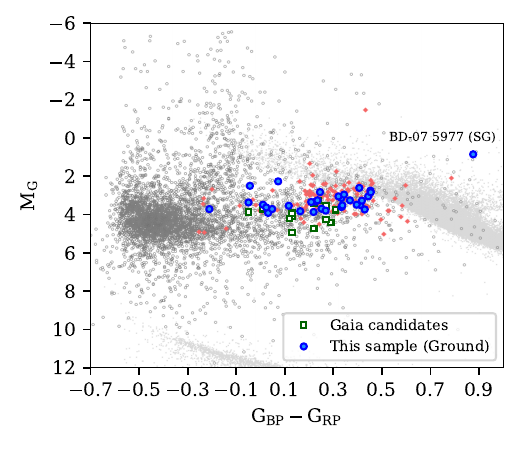}
  \includegraphics[width=8.8cm, height=7.5cm]{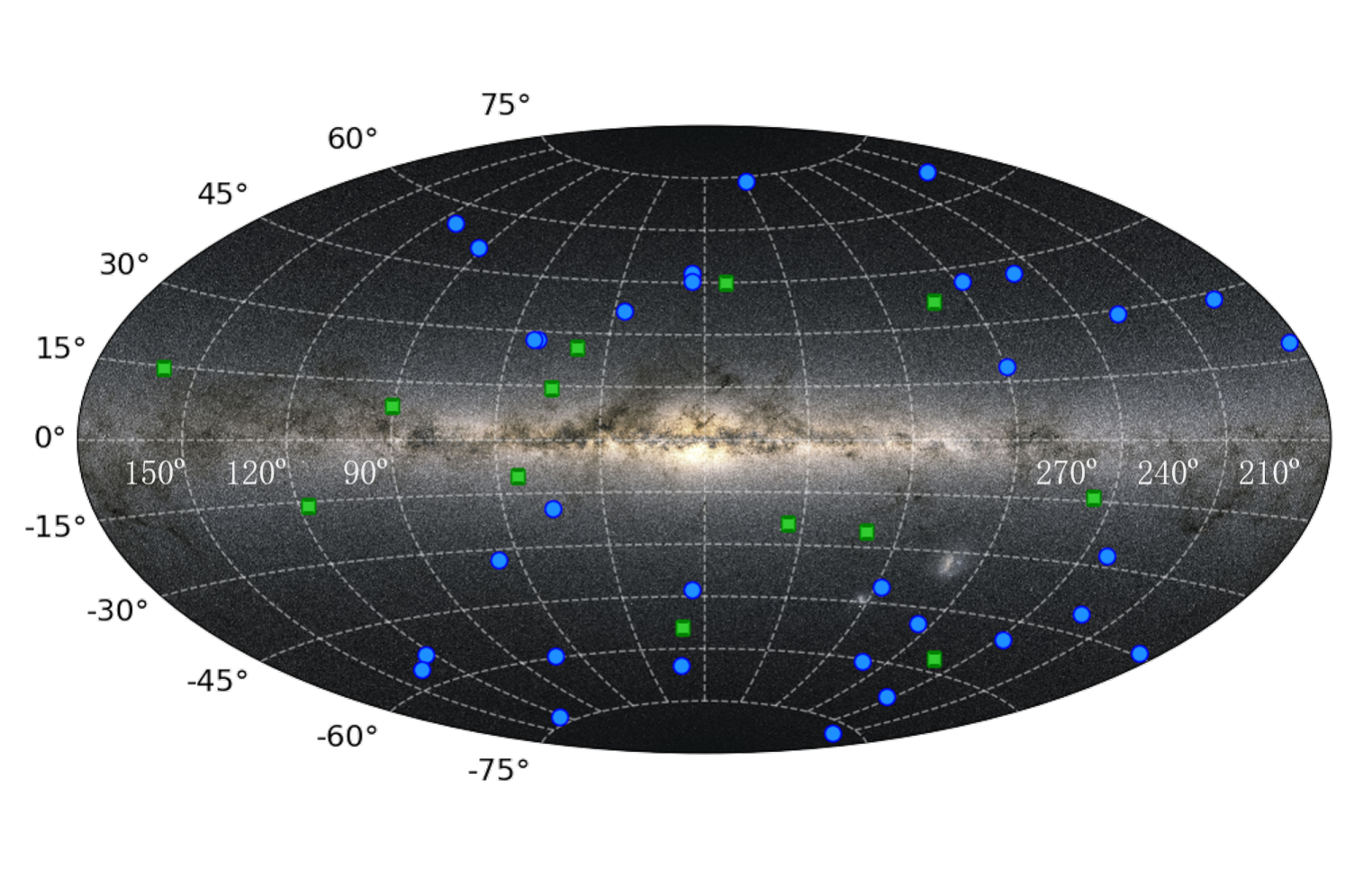}
      \caption{\textit{Left}: The color-magnitude diagram (CMD) presented by \citet{Pelisoli2020A&A} for their composite hot subdwarf sample is shown as dimmed red diamonds, representing systems whose light curves and rotation periods were studied by the authors. We have included the wide hot sdB binary Ground-based sample with fully solved orbital parameters as blue-filled circles. The Gaia-based sample is shown as green-filled squares. Interstellar extinctions for both, Ground- and Gaia-based samples, are not corrected. Stars catalogued by \citet{Geier2020A&A} as hot subdwarfs are represented as dark gray open circles. Main-sequence stars and stars in other evolutionary stages within the region by \citet[sample C,][]{Lindengren2018A&A}{}{} are shown as pale gray dots. \newline \textit{Right}: An Aitoff projection displaying the Galactic coordinates of the wide hot sdB sample. Galactic longitude increases from the centre toward the left, with increments of 30º per tick. Same color-coded wide sdB systems.}
\label{fig:21-Color_Galac}
\end{figure*}

So far, the sample of wide hot subdwarf binaries contains 32 systems whose orbits have been fully solved.
Their Galactic locations and color-magnitude diagram (CMD) are shown in Fig.\,\ref{fig:21-Color_Galac}. The Gaia G absolute magnitude is derived from the Gaia apparent magnitude (g) listed in Gaia DR3 according to:
\begin{equation}
    \label{Gabs}
        M_{G}= g + 5 \cdot \log_{10}(\pi/1000\,\text{mas}) + 5
\end{equation}

In the CMD, the sample occupies the extended region populated by hot subdwarf binary systems, located in redder colors than the EHB overdensity of single hot subdwarfs.
However, one system in the sample is significantly shifted: BD-07 5977. The cool companion of this system was previously studied and reported as a subgiant star \citep{Vos2017A&A}.

\subsection{Gaia sample}
\label{sub:gaia}

The non-single star catalogue from Gaia DR3 \citep[NSS\footnote{\url{https://cdsarc.cds.unistra.fr/viz-bin/cat/I/357}},][]{Gaia2023A&A} gathers the sources whose observations are considered as non-constant in one of the three channels of observation capabilities from the Gaia probe: astrometric, spectroscopic, and photometric.
This leads to different approaches and binary orbit models to achieve their orbital parameters\footnote{\url{https://gea.esac.esa.int/archive/documentation/GDR3/Data_analysis/chap_cu4nss/}}.

The data release 3 catalogues\footnote{\url{https://cdsarc.u-strasbg.fr/viz-bin/cat/J/A+A/662/A40}} by \citet{Culpan2022A&A} contain 6\,616 hot subdwarfs with spectroscopic confirmation and a further 61\,585 hot subdwarf candidates based on Gaia eDR3. They include multi-band photometry and Gaia eDR3 astrometry, as well as classifications based on spectroscopy and colors.

The cross-matching using TOPCAT\footnote{\url{https://www.star.bris.ac.uk/~mbt/topcat/}}\citep{Taylor2005ASPCS} between both, the NSS and \citet{Culpan2022A&A} catalogues, yields the coincidence of 13 sources. 
Fundamental data of the sources and their observations are listed in Table \ref{21-Obs} of the appendix, along with the wide sdB binary systems. 
Likewise, their CMDs are shown in Fig.\,\ref{fig:21-Color_Galac} (left panel), along with their Galactic coordinates (right panel).

\section{Orbital parameters}
\label{sec:Orbital}

\begin{table}
\caption{Orbital parameters from Gaia-based solutions (NSS) for the wide hot subdwarfs Gaia candidates: orbital periods (P) and eccentricity (e), including their standard errors ($\sigma$). The last field (NSS sol.) accounts for the data Gaia channel and approach from which they were determined. See text in Sect.\,\ref{sub:gaia} and especially \ref{sub:OP Gaia} for further details.}
\label{tb:33-OP_Gaia}
\centering
\renewcommand{\arraystretch}{1.5}
\begin{tabular}{l@{\hskip 0.078in}cccc@{\hskip 0.048in}c}
\hline
\hline
System Gaia DR3   & P(d) & {\tiny $\sigma_{\text{P}}$}(d) & e &  \tiny $\sigma_{\text{e}}$ &   NSS sol.  \\
\hline
\footnotesize 392046852459641472    & 884    & \tiny $\pm 25  $  & 0.497  & \tiny $\pm 0.090  $  &  \tiny Orbital    \\ 
\footnotesize 4850445797329363328   & 913    & \tiny $\pm 80  $  & 0.233  & \tiny $\pm 0.149 $   &  \tiny SB1         \\ 
\footnotesize 972725503164737152    & 854    & \tiny $\pm 32  $  & 0.269  & \tiny $\pm 0.098  $  &  \tiny Orbital     \\ 
\footnotesize 5579436712515286016   & 751    & \tiny $\pm 8   $  & 0.022  & \tiny $\pm 0.054  $  &  \tiny Orbital     \\ 
\footnotesize 3540092300847749760   & 617    & \tiny $\pm 5   $  & 0.087  & \tiny $\pm 0.068  $  &  \tiny Orbital     \\ 
\footnotesize 6335746093599431296   & 718    & \tiny $\pm 41  $  & 0.360  & \tiny $\pm 0.234  $  &  \tiny Orbital     \\ 
\footnotesize 4550114402362108416   & 821    & \tiny $\pm 43  $  & 0.152  & \tiny $\pm 0.040  $  &  \tiny Orbital     \\ 
\footnotesize 6359368722966483840   & 740    & \tiny $\pm 16  $  & 0.231  & \tiny $\pm 0.080  $  &  \tiny Orbital     \\ 
\footnotesize 4522995326025050496   & 978	 & \tiny $\pm 188 $  & 0.151  & \tiny $\pm 0.183  $  &  \tiny Orbital     \\ 
\footnotesize 6631822855308840320   & 634	 & \tiny $\pm 14  $  & 0.425  & \tiny $\pm 0.145  $  &  \tiny Orbital     \\ 
\footnotesize 1806581377789928448   & 847	 & \tiny $\pm 57  $  & 0.207  & \tiny $\pm 0.059  $  &  \tiny Orbital     \\ 
\footnotesize 2183496162809214464   & 1138   & \tiny $\pm 256 $  & 0.067  & \tiny $\pm 0.131  $  &  \tiny Orbital     \\ 
\footnotesize 6587335519633433472   & 828	 & \tiny $\pm 29  $  & 0.292  & \tiny $\pm 0.067  $  &  \tiny Orbital     \\ 
\hline
\end{tabular}
\label{tb:33-OP_Gaia}
\end{table}

\subsection{Radial velocities}
\label{sub:RV}

The radial velocities of the cool companion can be determined using cross-correlation (CC). The wavelength intervals used avoid telluric lines and those from the hot companion. They represent a good compromise between the high signal-to-noise ratio (SNR) and the low flux contribution of the hot component. 
Then, the lines of the cool companion within these intervals are cross-correlated against a high-resolution synthetic template with atmospheric parameters of a similar photometric class to the companion. Before the CC, the synthetic spectrum is convolved to reproduce the line broadening profile of the cool companion star due to the rotational velocity.

The masks utilised for the HERMES spectra are provided by its hermesVR pipeline\footnote{\url{http://hermes-as.oma.be/manuals/cookbook5.0.pdf}}. Meanwhile, for UVES spectra, Kurucz LTE synthetic templates were used \citep{Kurucz1979ApJS}.

The RV error intervals are obtained by performing a Monte Carlo (MC) simulation for each spectrum. 
Firstly, Gaussian noise is added to the spectrum, based on the continuum noise level in the used wavelength ranges. Then, the synthetic spectrum is used in a new cross-correlation, and the final error is calculated from the standard deviation of the RV results from 500 simulations.

The hot sdB stars pose a greater challenge for determining their RVs due to the limited number of available lines. In many cases, only the not blended \ion{He}{i} $\lambda$ 5876 \AA\, line can be used. However, some systems show a few sharp metal lines originating from the sdB companion. If that is the case, the intervals containing those lines are used. 

For the hot sdB stars, we employed a custom high-resolution template spectrum. The templates were generated using the {\sc XTgrid} code \citep{Nemeth2012MNRAS,Nemeth2019ASPCS}. 
The code uses a wavelength space direct spectral decomposition \citep{Simon1994A&A} to produce synthetic composite spectra through a linear combination of non-LTE ({\sc Tlusty}, \citealt{Hubeny2017}) and LTE ({\sc Atlas}, \citealt{Bohlin2017AJ}) atmosphere
models. The process involves an iterative fitting routine to the observed spectrum to determine the main parameters of both the hot and cool companions.
For obtaining basic templates efficiently and without being time-consuming, initial rough fits to the observed spectra are used as templates for cross-correlation.
This approach works, as for the CC, the exact line shape is less important than the presence of the respective lines. The RV error intervals are determined using the MC simulations as described above.

In Table \ref{tb:31-CCintervals} of the appendix, we indicate the intervals and/or the specific lines used for the CC of the MS and sdB companions.

\subsection{Keplerian fits and orbital parameters}
\label{sub:Kpfits}

The orbital parameters were obtained by simultaneously fitting Keplerian orbits to RVs of the companions \citep{Hilditch2001} with eight free parameters: orbital period ($P$), time of periastron ($T_0$), eccentricity ($e$), angle of periastron ($\omega$), two semi-amplitudes ($K_1$ and $K_2$) and two systemic velocities ($\gamma_1$ and $\gamma_2$). The systemic velocities of both components are considered as separate parameters, as the difference in the surface gravity between the components can cause a line-shift (gravitational redshift, see e.g. \citealt{Vos2012A&A, Vos2013A&A}). 

The error intervals on the orbital parameters are calculated using an MC approach, where in each iteration, each single RV measurement is randomly sampled using a normal distribution centred on its best value and a standard deviation equal to its standard error. The final orbital parameters are the mean values of 1000 iterations, while their standard deviations are used as error intervals.

\begin{figure}
   \centering
   \includegraphics[height=6.5cm,width=9cm]{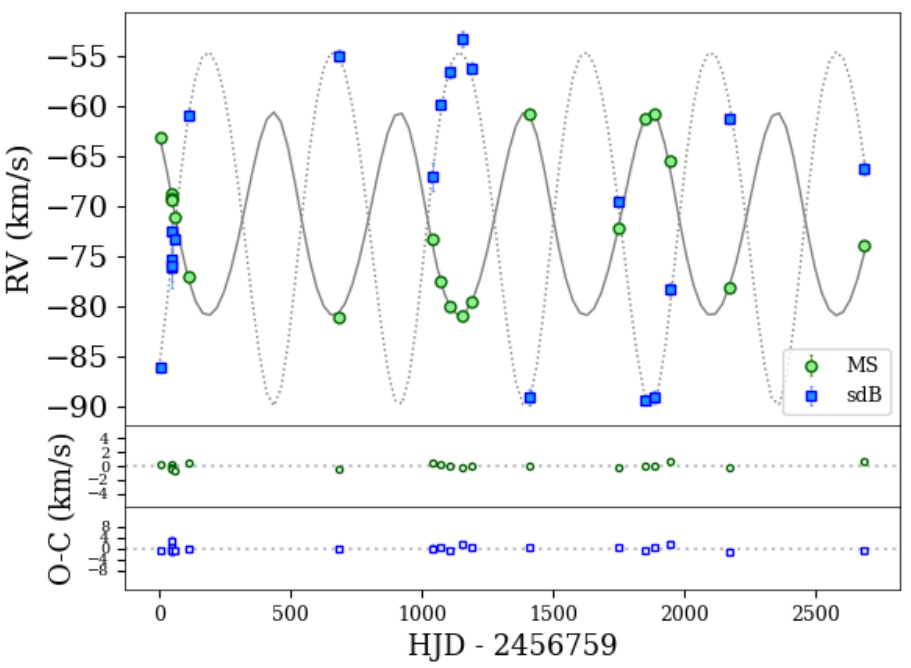}
      \caption{Radial velocity curves and residuals for PG1514+034. The RVs of the cool companion are plotted as green-filled circles, while those of the sdB companion are shown as blue-filled squares. RV error bars (1$\sigma$), obtained through MC simulations, are included. The best-fit Keplerian orbits, simultaneously fitted for both the MS and sdB companions, are represented by a solid line for the cool companion and a dotted line for the hot one.}
         \label{32-Kpfit}
   \end{figure}

Fig.\,\ref{32-Kpfit} shows an example of RVs and the simultaneously Keplerian orbits fitted to the companions of the PG1514+034 system.
Table \ref{tb:31-orbParam} of the appendix presents the orbital parameters obtained from the corresponding solutions for the sample of wide sdB binaries. 

\subsection{Gaia-based orbital solutions}
\label{sub:OP Gaia}

The orbital periods and eccentricities of the wide sdB systems whose orbital solutions were determined by Gaia are shown in Table \ref{tb:33-OP_Gaia}. The data are obtained from the astrometric and photometric channels of Gaia. Gaia orbital solutions of the sample, but for one system, are provided by the model for an astrometric binary source compatible with Campbell orbital elements \citep{Halbwachs2023A&A}. The so-called elements involve the semi-major axis of the orbit described by the photocentre, $a_0$, the inclination of the orbital plane with respect to the sky, $i$, the position angle of the ascending node, $\Omega$, and the periastron longitude measured from the ascending node, $\omega$. The measurement of the position of the photocentre relative to the barycentre of the system is related to these elements by a completely nonlinear relationship. The solutions of these elements are obtained by applying the formulas provided by \citet{Binnendijk1960book} or \citet{Halbwachs2023A&A}.

The $RA$ and $Dec$ coordinates and parallaxes (distances) of the Ground-based sample are provided from Gaia DR3. In contrast, for the Gaia sample, the solutions are from the NSS, since they are expected to supersede the original ones in Gaia DR3 \citep{Gaia2023A&A}. Hence, they are used in Table \ref{tb:33-OP_Gaia}. 
A detailed description of this solution approach and its constraints is included in the documentation provided on the Gaia-ESA Web\footnote{\url{https://gea.esac.esa.int/archive/documentation/GDR3/Data_analysis/chap_cu4nss/sec_cu4nss_astrobin/ssec_cu4nss_astrobin_orbital.html}}.

Orbital parameters for the Gaia DR3 system 4850445797329363328 are derived from its single-lined spectroscopic data (SB1) from the Gaia spectroscopic channel. Following methodology similar to that described in the previous subsection in relation to the SB2 spectra of the wide sdB sample, RVs are fitted using a Keplerian orbit to determine the key orbital parameters of the observed component \citep{Gosset2024A&A}: orbital period ($P$), time of periastron passage ($T_0$), eccentricity ($e$), argument of periastron ($\omega$), RV semi-amplitude ($K$), and systemic radial velocity ($\gamma$). The RV curve is fitted using a general Keplerian function via a least-squares optimization procedure. The quality of the fit is assessed using the reduced chi-squared ($\chi^2$) statistic. A detailed description of the fitting process in SB1 and its constraints is available in the Gaia-ESA documentation\footnote{\url{https://gea.esac.esa.int/archive/documentation/GDR3/Data_analysis/chap_cu4nss/sec_cu4nss_spectroSB1/ssec_cu4nss_spectroSB1_model.html}}.

\section{Atmospheric parameters of the cool companions}
\label{sec:AP_MS}

We studied the composite master spectra of the cool companion stars to derive their atmospheric parameters. They were created by shifting the spectra from single observations to the rest velocities of the cool star lines. Then, they were merged, producing a master spectrum with an improved SNR compared to the individual observations. For the analysis of the atmospheric parameters, we used the redward wavelength interval of 6000-6260 \AA\, as a compromise between a high SNR and the contribution of the cool companion.  
The interval was trimmed and normalized using a polynomial fit to manually selected continuum points.

Spectroscopic analysis was performed using the Grid Search in Stellar Parameters code \citep[][\gssp]{Tkachenko2015A&A}. \gssp utilizes a method based on atmospheric models and spectrum synthesis, comparing observations with each theoretical spectrum in the grid.
To generate synthetic spectra, this study uses the SynthV LTE-based radiative transfer code \citep{Tsymbal1996} and a grid of LTE atmosphere models precomputed with the {\sc LLmodels} code \citep{Shulyak2004A&A}. Furthermore, it uses an extended grid (\met $\leq$ -1.0 dex) of LTE Kurucz atmosphere models \citep{Kurucz1979ApJS} to work with the most metal-poor systems of the set.

\gssp simultaneously optimizes seven stellar parameters: effective temperature (\teff), metallicity ([M/H]), surface gravity (\logg), projected rotational velocity (\vsin), macro-turbulent velocity (\vmac), micro-turbulent velocity (\vmic), and the dilution factor ($F_{\rm MS}/F_{\rm Total}$) of the cool companion. A grid of theoretical spectra was built from all possible combinations of these parameters, and \gssp compares them to the observed normalized spectrum. A merit function $\chi^2$  is used to evaluate the match between the synthetic spectra and the corresponding MS master spectra, identifying the set of best-fit parameters.

The dilution factor was treated as a wavelength-independent parameter within the selected short wavelength interval. \citet{Vos2018MNRAS} compared this approach with other spectral analysis methods and found that using a fixed dilution factor does not compromise accuracy when applied to short-wavelength ranges.

The values of \vmic and \vmac were not freely refined but were fixed in \gssp because the SNR of the master spectra is not high enough to avoid degeneracy and determine these parameters accurately.
For these parameters, we adopted the iterative procedure described in \citet{Molina2022A&A}, which uses the calibrated relations for \vmic from \citet{Bruntt2010MNRAS} and for \vmac from \citet{Doyle2014MNRAS}. These relations are based on spectroscopic and asteroseismic data of MS field stars.

The initial setup for the parameter search covers a wide range with a large step size to ensure that the global minimum is found while minimizing computational cost. Based on the analysis of this first coarse grid, two consecutive setups with smaller step sizes are implemented. The parameters obtained from each iteration are used to refine the setup for the subsequent search.
The outcome of final parameters and their associated errors is determined from the last iteration by fitting a polynomial function to the reduced $\chi^2$ coefficients. The minimum of the polynomial defines the final parameter value, while the 1$\sigma$ cut-off values provide the error intervals.

\begin{figure*}
  \centering
  \includegraphics[width=13cm]{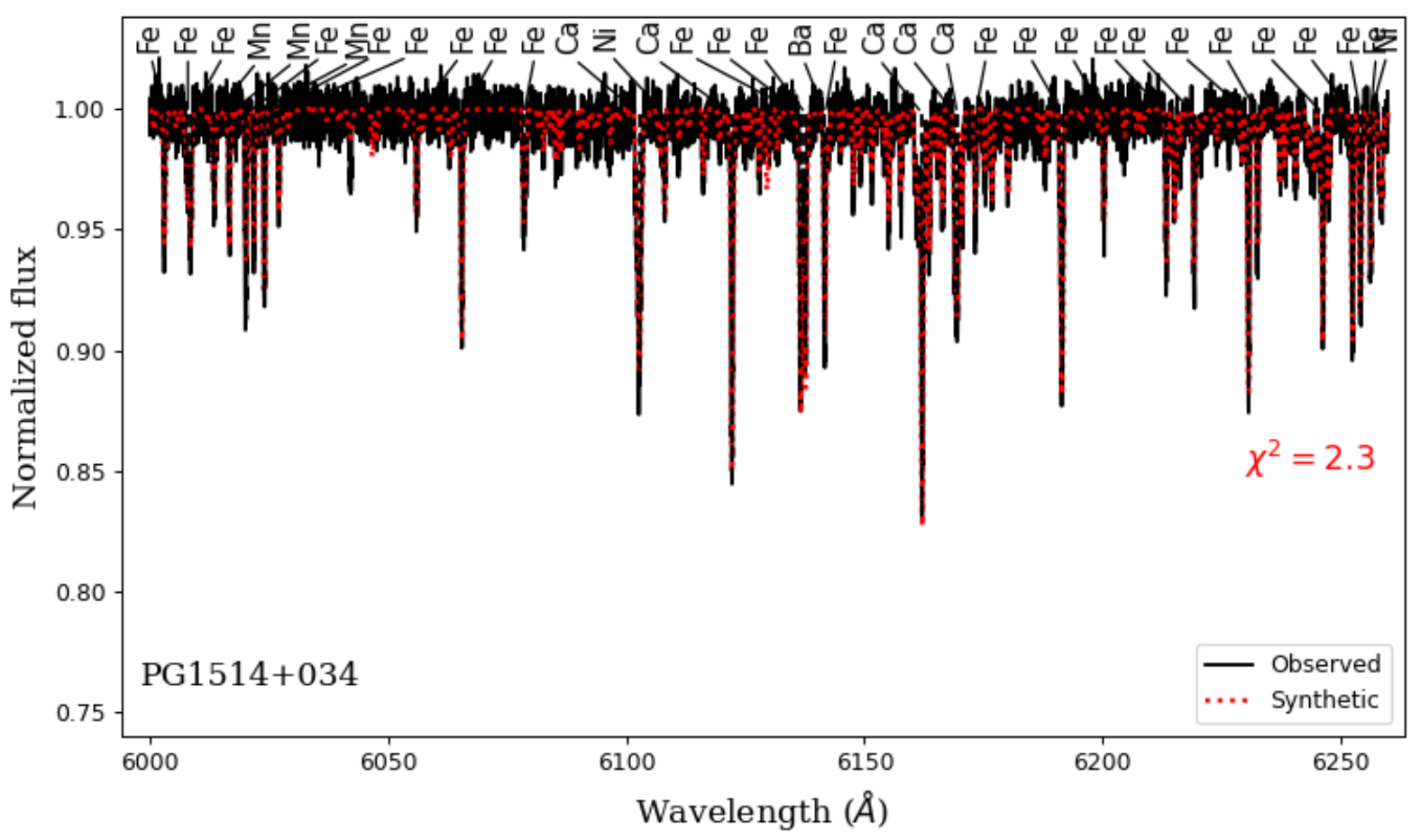}
      \caption{The observed normalized spectrum (black solid line) of PG1514+034 and the best-fitting \gssp model to the cool companion lines (red dotted) for the wavelength range of 6000-6260 \AA, used to determine the stellar atmospheric parameters of the cool companion.}
\label{fig:40-PA_PG1514+034}
\end{figure*}

\begin{figure}
  \centering
  \includegraphics[height=5.4cm,width=8.0cm]{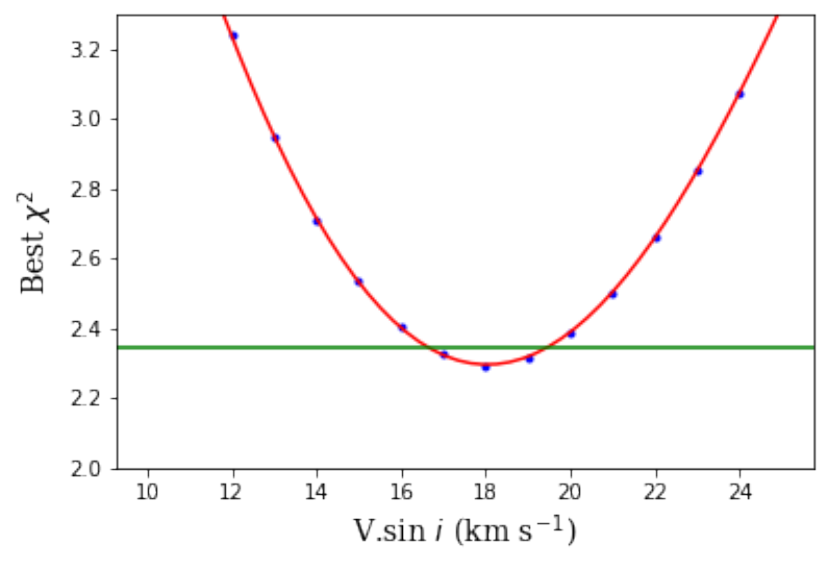}
      \caption{The process of obtaining \vsin and its error interval for the cool companion of PG1514+034. The fitting of the polynomial function (in red) to the reduced $\chi^2$ coefficients (dark blue points) yields the outcome of the atmospheric parameter as the minimum, while the 1$\sigma$ cut-off (green bar) provides its error interval.}
\label{fig:40-Vsini_PG1514+034}
\end{figure}

Fig.\,\ref{fig:40-PA_PG1514+034} shows the best-fitting \gssp synthetic model for the wavelength range of 6000-6260 \AA, applied to the normalized master spectrum of PG1514+034 (also catalogued as EGGR440). Fig.\,\ref{fig:40-Vsini_PG1514+034} illustrates the derivation of the atmospheric parameter results and their error intervals, in this example, \vsin of PG1514+034. The atmospheric parameters derived and their corresponding error intervals for the cool companions of the wide hot sdB binary sample are presented in Table \ref{GSSP} of the appendix.

\section{Observed properties}
\label{sec:Properties}

\subsection{Galactic orbits and kinematics}
\label{sub:Galactic}

\begin{figure*}
  \centering
  \includegraphics[width=7.4cm]{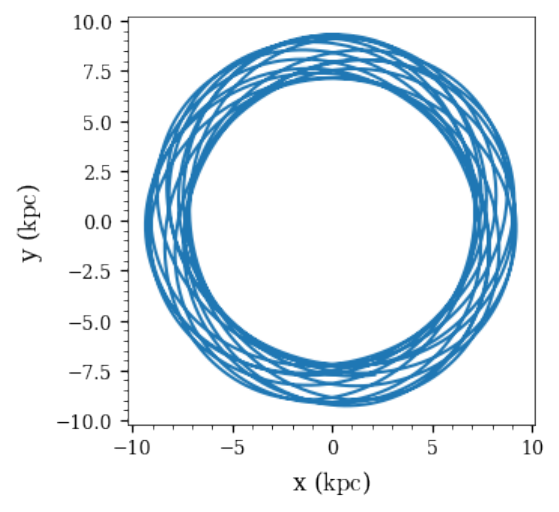}
  \includegraphics[width=7.10cm]{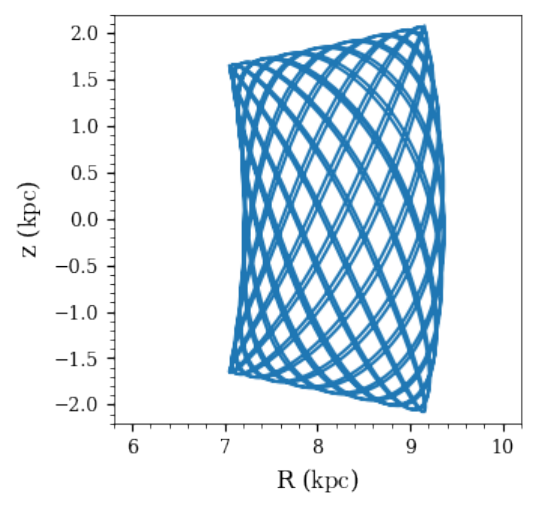}
      \caption{The Galactic orbit of PG1514+034 projected onto the Galactic plane (\textit{left}) and the vertical Galactic component versus the radius with respect to the Galactic center (\textit{right)}.}
\label{fig:51-org_PG1514}
\end{figure*}

The Galactic orbits of the wide hot subdwarf binaries are derived using parameters such as their coordinates (\textit{RA}, \textit{Dec}), distance, proper motions in the \textit{RA} (\textit{pmra}) and \textit{Dec} (\textit{pmdec}) directions, and systemic radial velocity ($\gamma$). As an example, the Fig.\,\ref{fig:51-org_PG1514} illustrates the derived Galactic orbit of the PG1514+034 system, computed using the Python package for Galactic dynamics, {\sc galpy}\footnote{\url{https://docs.galpy.org/}}.

\begin{figure*}
\centering
\includegraphics[width=12.5cm]{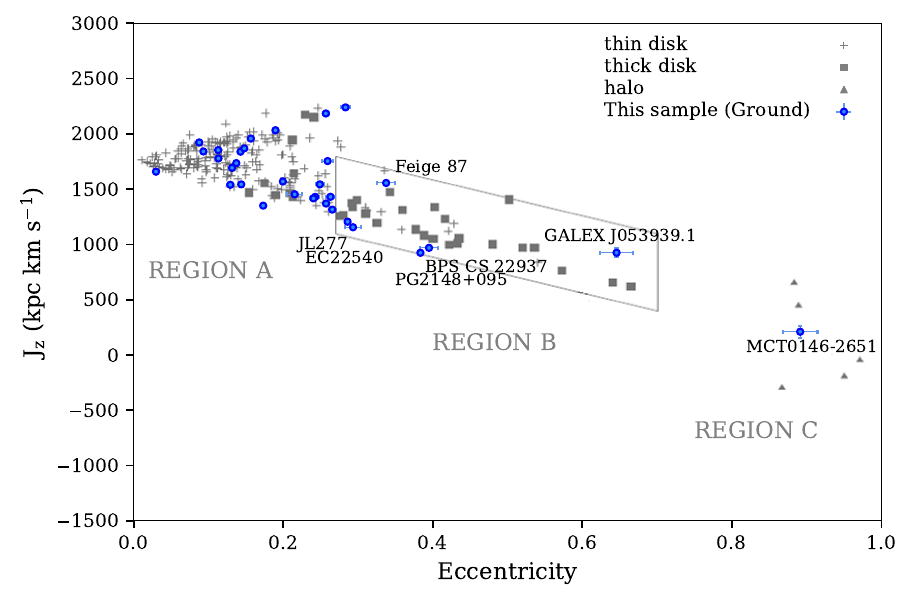}
  \caption{J$_Z$ versus Galactic orbit eccentricity for the calibration sample of main-sequence stars from \citet{Pauli2006A&A}. Region A is populated by thin disk stars, region B by thick disk stars, and region C by halo stars. The plot has been adapted to include the Galactic orbit parameters of the wide hot subdwarf sample. The wide sdB systems of the sample, which likely belong to the thick disk or the halo, are labelled. Error bars are included.
  }
  \label{fig:51-GalaOrbits}
\end{figure*}

The diagram in Fig.\,\ref{fig:51-GalaOrbits} shows the $z$-component of the angular momentum (J$_Z$) against the eccentricity of the Galactic orbits for a calibration sample used by \citet{Pauli2006A&A}. The sample of wide hot sdB binaries has been incorporated into this diagram.
The calibration sample consists of 291 F- and G-type MS stars, whose Galactic component memberships were previously determined based on both kinematic and chemical criteria \citep{Edvardsson1993A&A,Fuhrmann1998A&A,Fuhrmann2004AN}. Three distinct regions (A, B, and C) are identified in the diagram, predominantly populated by thin-disk, thick-disk, and halo stars, respectively. The positions of the wide sdB binary systems within the diagram suggest their probable Galactic population membership. However, the wide sdB binary sample consists of systems in which at least one companion has undergone post-main-sequence evolution. Hence, it is an older and more metal-poor population, which could be associated with also more eccentric Galactic orbits and dynamically hotter kinematics \citep{Wu2021MNRAS,Nogueras2024A&A}.

The eccentricity and J$_Z$ of MCT0146-2651 (also known as SB744) strongly suggest its affiliation with the halo population. Recently studied by \citet{Nemeth2021A&A}, the chemical analysis revealed a low metallicity (\met = -1.09 dex), providing additional support, along with their kinematic parameters, for its likely halo membership.

\begin{figure*}
\centering
\includegraphics[width=11.5cm]{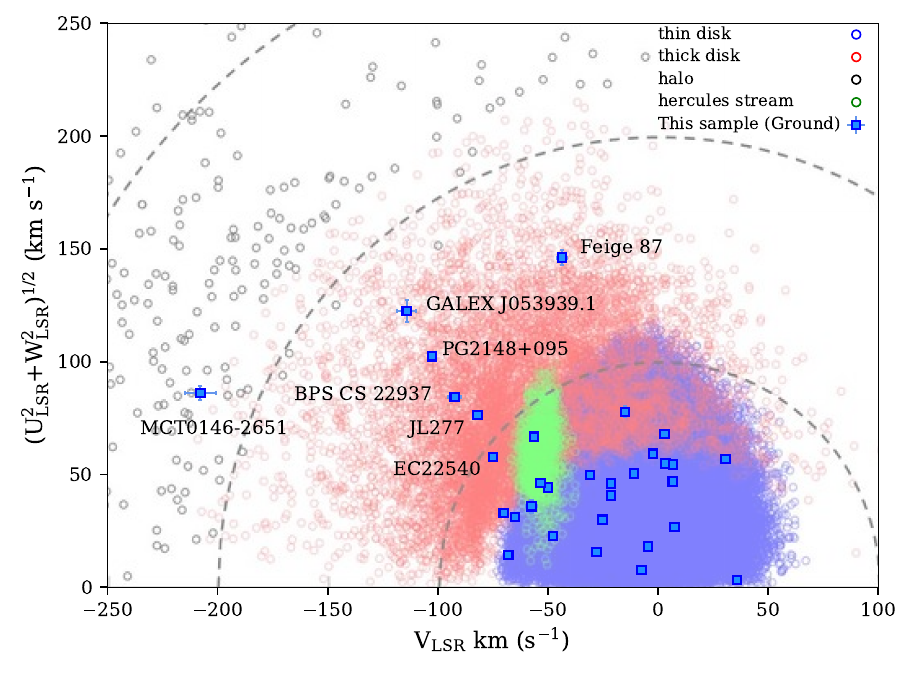}
  \caption{The Toomre diagram of the calibration sample by \citet{Chen2021ApJ} for the different Galactic components, attending to kinematics and stellar age criteria (see their paper for thresholds info). The calibration sample contains main-sequence turn-off and subgiants stars from the solar neighbourhood to $\approx$ 1500 pc. The diagram has been adapted (in dimmed colors) to overlap the wide hot subdwarf binary sample. The sdB systems, which are more likely to belong to the thick disk or the halo, are labelled. Included error bars.}
  \label{fig:51-Toomre}
\end{figure*}

For the Galactic kinematic study, the velocity components (U, V, W) of the sdB sample were determined using the criteria and calculations outlined in \citet{Johnson1987AJ}. These velocities were corrected for solar motion relative to the local standard of rest (LSR) following the criteria of \citet{Coskunoglu2011MNRAS}. The results are presented in a Toomre diagram in Fig.\,\ref{fig:51-Toomre}.

The diagram includes a calibration sample from \citet{Chen2021ApJ} for the different Galactic components, incorporating data from the {\sc LAMOST MSTO-SG} and Gaia DR2 catalogs. The classification of stars into various Galactic components was determined based on kinematic, chemical, and evolutionary considerations, as described in their study.
Nevertheless, as generally accepted and indicated, stars with $V_{tot}= (U_{LSR}^2+V_{LSR}^2+W_{LSR}^2)^{1/2}  < 50$ km s$^{-1}$ predominantly, but not exclusively, belong to the thin disk, while those with moderate velocities $V_{tot} \approx 70-180$ km s$^{-1}$, are more typically associated with the thick disk \citep[see f.i.][]{Adibekyan2013A&A,Bensby2014A&A}{}{}.
Stars linked to the Hercules stream typically meet the criteria of $V_{LSR} \approx -50 \pm 9$ km s$^{-1}$ and $(U_{LSR}^2+W_{LSR}^2)^{1/2} \approx 50-70$ km s$^{-1}$ \citep[e.g.][]{Famaey2005A&A,Bensby2007ApJ}{}{}.
Finally, halo stars are typically characterized by $V_{tot} > 220$ km s$^{-1}$ \citep{Bonaca2017ApJ}.

Most sdB systems are associated with the thin disk. However, consistent with Fig.\,\ref{fig:51-GalaOrbits}, some systems are identified as probable candidates for the old thin disk or thick disk: BPS CS 22937, PG2148+095, Feige 87, EC 22540-3324, and JL277. GALEX J053939.1-283329 is a strong candidate for the thick disk, and MCT0146-2651 is identified as a halo system, in agreement with previous findings involving their Galactic orbits.

\subsection{Parameter relations and comparison with binary evolution models}
\label{sub:models}

The determination of the complete set of orbital and cool companion atmospheric parameters for the current sample of wide hot sdB binaries represents an effort that spanned more than a decade of observations. Work is still underway to increase the number of systems.
Nevertheless, the sample obtained so far allows us to establish relationships between orbital and atmospheric parameters and to provide strong constraints on current theoretical frameworks.

\begin{figure*}
  \centering
  \includegraphics[width=8.5cm]{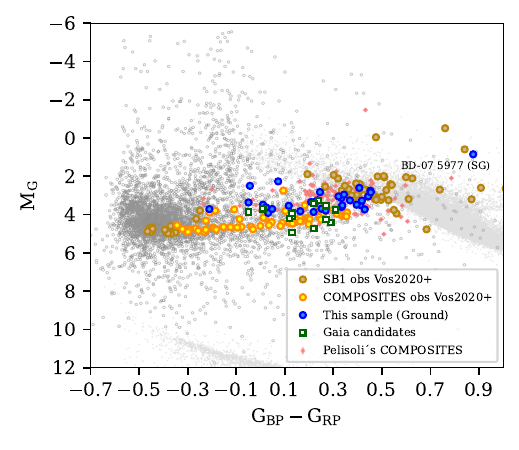}
  \includegraphics[width=8.5cm]{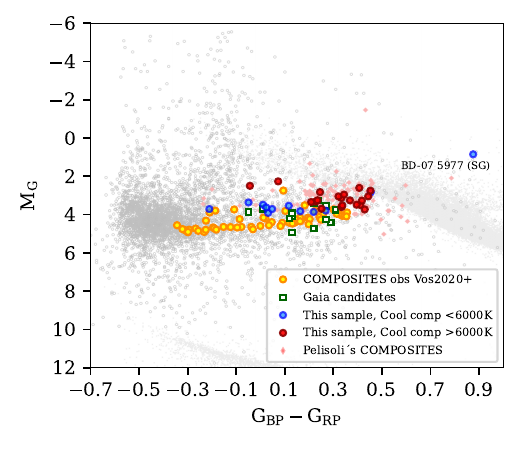}
      \caption{Same CMD diagram than in Fig.\,\ref{fig:21-Color_Galac} but including the simulations by \citet{Vos2020A&A}. \newline
      \textit{Left panel:} Brown filled circles indicating sdB SB1 observable simulations and orange empty ones for observable sdB composites, according to the visibility criteria of the theoretical work.\newline
      \textit{Right panel:} Only including simulations of observable sdB composites as orange empty circles. The systems in the wide sdB sample with cool companions whose \teff is higher than 6000 K are marked with red filled circles.
      }
\label{fig:52-CMD_Vos}
\end{figure*}

The most detailed long-period sdB population study to date, by \citet{Vos2020A&A}, used a combination of detailed stellar evolution simulations based on Modules for Experiments in Stellar Astrophysics \citep[MESA,][]{Paxton2011ApJS,Paxton2013ApJS,Paxton2015ApJS,Paxton2018ApJS,Paxton2019ApJS}, the observationally calibrated Besan\c{c}on Galactic model \citep{Robin2003}, model for synthetic observations and a standard model of binary interactions. The authors demonstrated that the $P-q$ distribution of the wide sdBs is partly determined by the metallicity history of the Galaxy and were able to reproduce the observed formation rates, metallicity, period, and mass-ratio distributions. Fig.\,\ref{fig:52-CMD_Vos}, shows the CMD diagram of Fig.\,\ref{fig:21-Color_Galac}, but this time including simulations from \citet{Vos2020A&A}. 

We note that a significant number of spectroscopic wide sdB binaries occupy a region predicted by the \citet{Vos2020A&A} theoretical model to contain SB1 systems, and are thus offset from the predictions for composite sdB simulations (left Panel). The simulated sdB systems show $ BP-RP$ colors redder than 0.3 and are dimmer by up to $\approx$1 G-band magnitude than the spectroscopic sample. Indeed, a similar conclusion can be drawn for the majority of the composite sdB sample, as reported by \citet{Pelisoli2020A&A}, which is included in the CMD.
A close analysis of the systems showing colors beyond 0.3, aside from the SG system BD-07 5977 \citep{Vos2017A&A}, indicates that they correspond mostly to those sdB containing cool companions with \teff higher than 6000 K (see right Panel of Fig.\,\ref{fig:52-CMD_Vos}), indicative of \citet{Vos2020A&A} overpredicting the number of the sdBs with the coolest companions. Hence, stars are likely to belong to the F- or earliest G-type spectral class at least. This issue will be addressed in a forthcoming work \citep{Molina2026paperII}.
Observational data from the wide sdB sample will likely help improve the visibility criteria in ongoing and future theoretical work.

\begin{figure*}
\centering
\includegraphics[width=9cm]{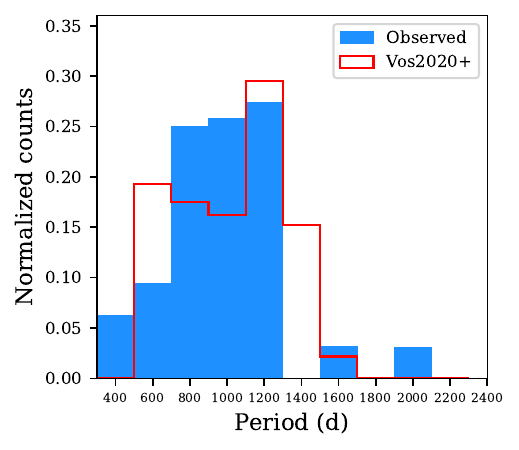}
\includegraphics[width=9cm]{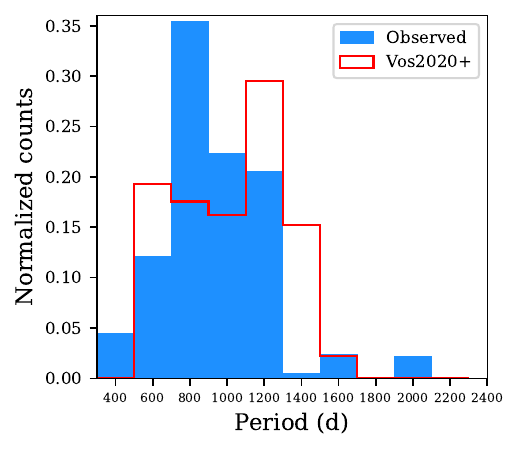}
  \caption{The distribution histogram of orbital periods and the predicted period distributions from the MESA simulations, according to the Galactic model by \citet{Vos2020A&A}.\newline
  \textit{Left}: Distribution histogram of the wide sdB binary systems sample not including Gaia wide sdB candidates.\newline 
   \textit{Right}: Distribution histogram including both, the wide sdB binary systems sample and Gaia wide sdB candidates.
  }
  \label{fig:52-PVos}
\end{figure*}

The study of the radial velocity curves provides insight into the orbital period distribution of the wide hot sdB sample.
The orbital period distribution has been studied since the theoretical work by \citet{Vos2020A&A}. Fig.\,\ref{fig:52-PVos} shows the histogram of predicted orbital period distributions from the simulations, based on the visibility criteria, in agreement with the Galactic model.

To construct this histogram, the same method used by \citet{Vos2013A&A} was employed, treating each observation as a normal distribution with the error as the standard deviation. Subsequently, 1,000 orbital periods per system were randomly generated within the uncertainty interval. The resulting 32,000 points (from 32 systems) were binned into intervals of 200 days. 

The inclusion of wide sdB Gaia candidates in the orbital period distribution is notably disruptive in the comparison between theoretical and observational distributions. This discrepancy could arise from an observational bias affecting the Gaia sample, since the selection criteria for the Gaia binaries are not identical to those used for the spectroscopic sample. The reported orbital periods tend to cluster in the 700–900-day range, with significant uncertainties in some systems. Alternatively, these Gaia candidates might indicate the existence of a new population of wide sdB systems that has not yet been accounted for by the spectroscopically confirmed sample.
Their positions in the above-mentioned CMD diagrams indicate fainter absolute magnitudes than populations of the current wide sdB or Pelisoli's samples.

\begin{figure*}
\centering
\includegraphics[width=8cm]{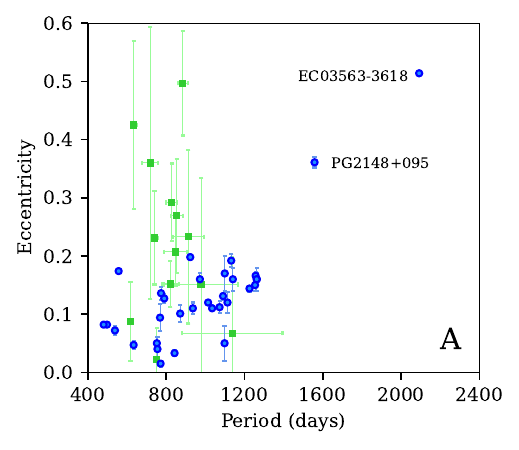}
\includegraphics[width=8cm]{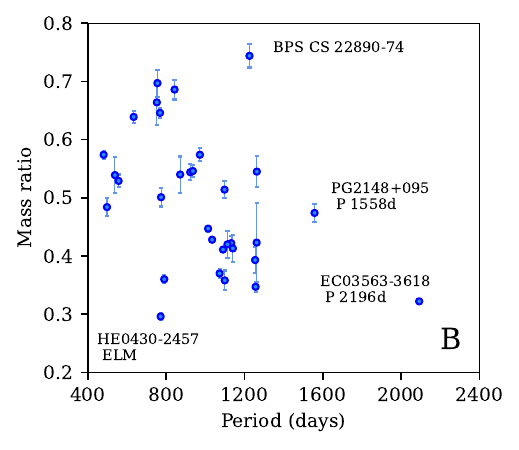}
\includegraphics[width=8cm]{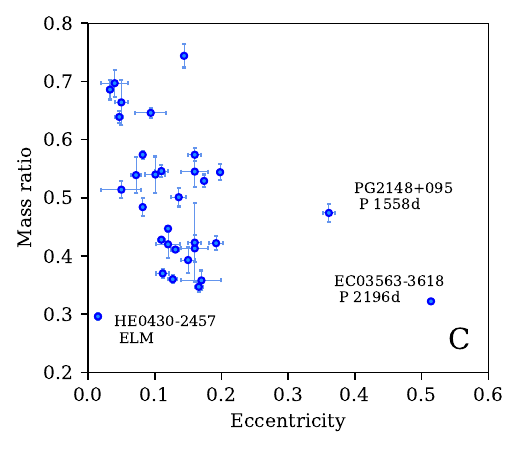}
\includegraphics[width=8cm]{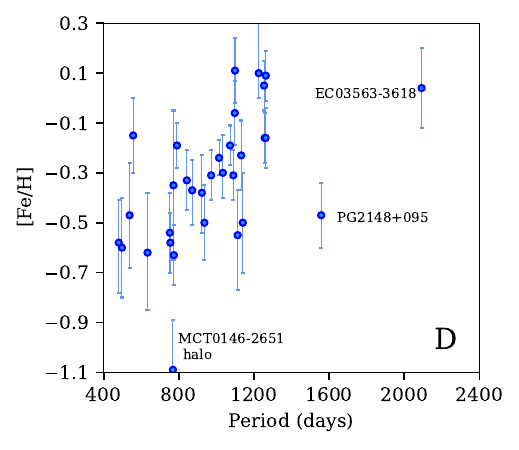}
\caption{Correlations between orbital parameters derived from the RV curves of the wide sdB binary systems. Error bars are included. 
 \newline \textit{Panel A:} Diagram of the orbital period-eccentricity correlation.
 \textit{Panel B:} Diagram of the orbital period-mass ratio correlation.
 \textit{Panel C:} Diagram of the eccentricity-mass ratio correlation.
 \textit{Panel D:} Diagram of the orbital period-metallicity correlation.
 }
  \label{fig:52-Group}
\end{figure*}

The eccentricity of observed binary systems has long challenged theoretical models. Tidal forces are typically expected to lead to orbital circularization. However, observations suggest that eccentricity appears to increase with longer orbital periods (see Fig.\,\ref{fig:52-Group}, Panel A). 
\citet{Vos2015A&A} proposed a theoretical framework to explain this feature, based on the combination of two eccentricity-pumping mechanisms: phase-dependent RLOF and the formation of a circumbinary disk (CB), which can increase the final eccentricities of the systems. However, while the model successfully accounts for most systems, it is not a population model and does not fully reproduce the observed distribution of systems with the longest and shortest orbital periods.

A closer look at Panel A of Fig.\,\ref{fig:52-Group}, leads us to conclude that, excluding the two systems with unexpectedly long periods (and the largest eccentricities), the remaining systems do not exhibit such a strong correlation, as initially indicated by the first solved systems more than a decade ago. In our view, this suggests that additional complex factors might be involved in defining the final eccentricity of the systems. 

Similarly to the period distribution, the incorporation of the wide sdB Gaia candidates is disruptive in the diagram. This inconsistency may reflect an observational bias between the Gaia and spectroscopic samples, as previously noted regarding the CMD diagram and the orbital period issues. Nonetheless, it should be observed that the Gaia eccentricity measurements are subject to large uncertainties. On the other hand, a similar discrepancy between the eccentricity obtained by Gaia and the ground-based RV measurements of the binary system BD+20 5391 has been observed by \citet{Kurpas2025A&A}. The authors concluded that Gaia was likely overestimating the eccentricity of the binary system.

\begin{figure}
\centering
\includegraphics[width=9cm]{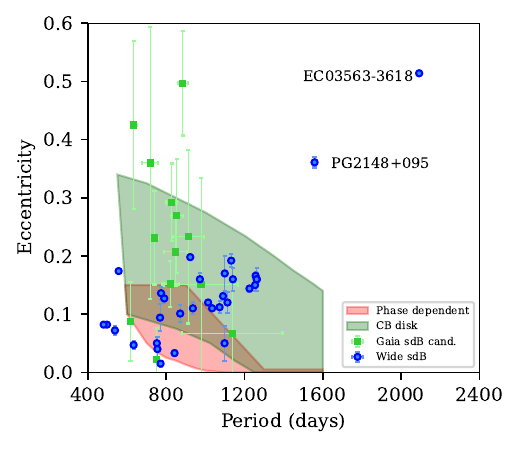}
  \caption{An adapted diagram of the orbital period-eccentricity correlation from \citet[see Fig. 9,][]{Vos2015A&A}. The red-shaded area represents the region explained by RLOF models, while the green-shaded area corresponds to a model incorporating a CB disk formation. The observed systems (blue dots) are plotted with error bars.}
  \label{fig:52-P_Ecc_Vos}
\end{figure}

Fig.\,\ref{fig:52-P_Ecc_Vos} illustrates the agreement between the observations and the theoretical regions predicted by the eccentricity mechanisms from \citet{Vos2015A&A}. The Gaia candidate sample further reinforces the apparent discrepancy with the wide sdB sample obtained through spectroscopic methods, but in addition to theoretical models. 
This might indicate another part of the population detected by Gaia using astrometric methods. However, as previously noted, their large uncertainties do not allow us to reach strong conclusions.

The formation of a CB disk during the mass loss phase in these systems has not yet been supported by empirical evidence. 
However, this idea is further supported by the numerous detections and evidence of dusty disks around binary systems, such as eccentric post-AGB binaries \citep[see e.g.][]{Kluska2022A&A, OOmen2020A&A,OOmen2018A&A,vanWinckel2017IAUS}, post-RGB systems \citep{Sarkar2022IAUS, Kamath2015EAS, Montgomery2012PASP}, and post-common-envelope binaries such as the white dwarf binary NN Serpentis \citep{Hardy2016MNRAS} or the recently published involving a sample of post-common-envelope close hot subdwarf binaries \citep{Li2025MNRAS}.

The formation of dust and the associated dusty and debris discs in binary systems has also been addressed from the theoretical point of view \citep[see e.g.][]{Bermudez2024MNRAS,Gonzalez2024MNRAS, Sarkar2022ApJ, Thebault2021A&A}. Although the gas would be removed early on from a potential CB disk in wide sdB binaries due to photoevaporation, this mechanism is argued to be inefficient for the destruction of relatively small grains ($\geq 0.01 \,\mu m$), even in the presence of a strong radiation field \citep{Nanni2024A&A}.
Hence, dust grains could have survived the He core and shell-burning phase.
Partially supporting this idea, the recent work by \citet{Li2025MNRAS} argues for the indirect detection of long-lived circumstellar material around close hot subdwarf binaries as likely remnants of their past common-envelope phase. If remnant material from past CB discs is still linked to wide sdB binaries, contrary to the optically thick discs of Post-AGB or -RGB, a negligible excess in near-IR should be expected \citep[see e.g. transition disks detected in Post-AGB by][]{Kluska2022A&A}. 

Two systems from the wide sdB sample show a slight IR excess in the Longwave-IR WISE W4 band, BD-11 162 and BD-7 5977. They were studied with the ALMA radio telescope \citep{Molina2022A&A}. However, no remnants of a CB disk were detected. In the case of BD-7 5977, the IR excess may be attributed to the subgiant nature of the cool companion \citep{Vos2017A&A}.
Nevertheless, these results do not rule out the presence of a CB disk in the history of these systems.

According to the previously outlined arguments, the direct detection is complicated, and indirect mechanisms should be developed, as utilised by \citet{Li2025MNRAS}. 
Their analysis was based on the presence of the \ion{Ca}{ii} K line, which is linked to the system. This line has been widely used in various studies to trace both transient and stable absorption features from CSM and dusty disks around different types of systems \citep[see e.g.][]{Vanderbosch2020ApJ,Rebollido2018A&A,Iglesias2018MNRAS,Welsh2016PASP,Montgomery2012PASP}. This topic, regarding the current sample of wide sdB with fully solved orbital parameters, will be explored in a forthcoming paper \citep[]{Molina2026ECCa}. Confirming the presence of remnant CSM around the systems would suggest that the systems once formed a CB disk.

\begin{figure}
\centering
\includegraphics[width=9cm]{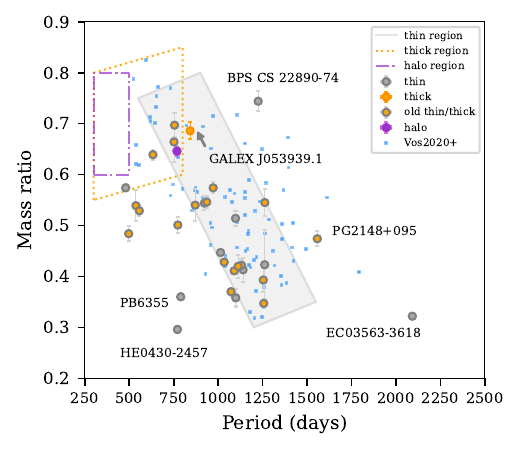}
  \caption{An adapted diagram of the orbital period-mass ratio correlation from \citet[see Fig. 7,][]{Vos2020A&A}. Simulations of wide sdB are shown as sky-colored filled squares. The regions where wide sdBs formed from different Galactic subpopulations are expected to be located are color-coded: thin-disk objects in gray, thick-disk objects in orange, and halo objects in purple. Similarly, the observed systems are color-coded according to their likely Galactic membership. Systems with unclear Galactic membership, thin or thick disk, due to the old population of wide sdB binaries are labelled as old thin/thick. See Sect.\,\ref{sub:Galactic} and \ref{sub:models} for a further explanation.}
  \label{fig:52-P_q_Model}
\end{figure}

The correlation between orbital period and mass ratio ($P-q$) in wide sdB binaries was studied by \citet{Vos2020A&A}. Their population study concluded that this correlation is strongly linked to the metallicity history of the Galaxy. Furthermore, the model predicts the final $P-q$ relation in the context of the Galactic membership of the systems. Fig.\,\ref{fig:52-Group} (Panel B) shows the $P-q$ observations. In Fig.\,\ref{fig:52-P_q_Model}, they are compared to the model of \citet[see their Fig. 7,][]{Vos2020A&A}. The sdB systems are color-coded based on their probable membership in the Galactic population, which we define using Galactic orbit eccentricity and the boundaries from the calibrated sample by \citet{Pauli2006A&A} in Fig.\,\ref{fig:51-GalaOrbits}, as well as the kinematic regions in Fig.\,\ref{fig:51-Toomre}. Systems with Galactic orbital eccentricity less than 0.15 were classified as strong Galactic thin disk candidates. Systems with Galactic orbital eccentricities between 0.15 and 0.45 were classified as a mixture of old thin disk and thick disk (i.e. belonging to either Region A or B). The system GALEX J053939.1-283329 in Region B, with an eccentricity higher than 0.45, consistent with its kinematics, was identified as a strong thick disk candidate. Finally, the system in Region C, MCT0146-2651, given its high Galactic orbital eccentricity and low J$_z$ (as well as its kinematics and low metallicity), is defined as a strong halo candidate.

Comparing the $P-q$ relation for the observed ground sample, classified by its Galactic membership, with the simulated sample from \citet{Vos2020A&A} in Fig.\,\ref{fig:52-P_q_Model}, we find both good agreement and new open questions. We can see that the bulk of the thin disk lies in the main part of the $P-q$ relation, overlapping with the simulated sample. The systems with low mass ratio ($<0.35$) and shorter orbital periods ($<1000\,\text{d})$, PB6355 and HE0430-2457, are consistent with being the extension of the \citet{Vos2020A&A} population towards younger and more massive, non-degenerately-igniting progenitors. The halo candidate system MCT0146-2651 is consistent with the metal-rich end of the Galactic halo, noting that the Besan\c{c}on model used in \citet{Vos2020A&A} assumes too narrow a metallicity range for the halo. The strong thick disk candidate GALEX J053939.1 is consistent with the metal-rich part of the Galactic thick disk. Finally, the region with mass ratios of $0.5$-$0.6$ and periods below $750\,\text{d}$, considered to be an unexplained second branch in \citet{Vos2020A&A}, can now be seen to mostly consist of old stars, offering a strong clue to its explanation in the population model. Systems PG 2148+095 and especially EC03563-3618 have periods that are very hard to explain by binary models and might be a product of triple stellar evolution \citep{Toonen16, Preece22}. Finally, clustering of binaries in several regions of the $P-q$ plane, as well as an excess of systems with $P$ close to $1250\,{\text{d}}$ are new emerging features that need further explanation.

The binary model by \citet{Vos2020A&A} also predicts a theoretical relationship between the final orbital periods and the metallicity of the systems. The Fig.\,\ref{fig:52-Group} (Panel D) shows the parameters derived from the observations, which are compared with the region predicted by the model. The general agreement is reasonably good. Some outliers are observed, including the two systems with the longest orbital periods (notably, by approximately a factor of two compared to the rest), EC03563-3618 and PG2148+095, as was also the case in the $P-Ecc$ correlation. MCT0146-2651 (or SB744), likely a halo member based on its Galactic orbit parameters, kinematics, and metallicity, may also be potentially deviating from the overall relation. At the same time, the second-branch systems in \citet{Vos2020A&A}, with orbital periods below $750\,\text{d}$ belong to the continuation of the same trend, again suggesting that they are part of the same population. 

\begin{figure}
\centering
\includegraphics[width=9cm]{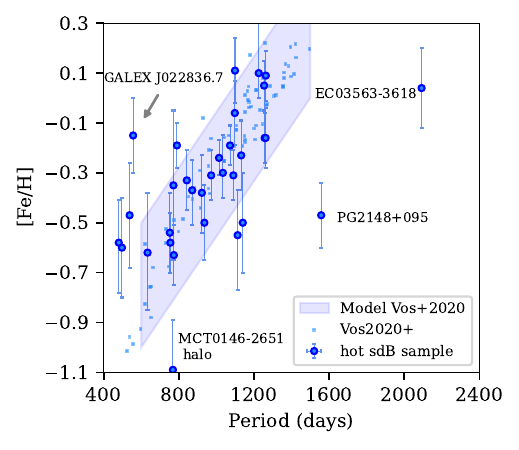}
  \caption{Orbital period versus metallicity for wide sdB binaries. The pale blue region represents the theoretical correlation predicted by \citet{Vos2020A&A}, based on a Galactic evolution model combined with a deterministic binary interaction model. Simulations of wide sdB are shown as sky-colored filled squares, wide sdB binaries sample as blue filled circles. Error bars are included.}
  \label{fig:52-P_Fe_Model}
\end{figure}

The sdB companion of MCT0146-2651 exhibits heavy-metal signatures (such as lead and fluorine), which led \citet{Nemeth2021A&A} to conclude that its formation probably resulted from a merger event in a hierarchical triple system. However, given that the system belongs to the binary population, the most likely explanation is that the heavy-metal signatures result from initial abundances and binary evolution.  Additionally, the system EC03563-3618 shows spectral disturbances, suggesting the presence of an additional companion or circumstellar material, possibly an unseen, very long-period cool dwarf. This is a matter for our next study that will outline the observed anomalies in the high-resolution spectra of this system \citep{Molina2026ECCa}. 

\section{Conclusions}
\label{sec:conclusions}
The orbital parameters of the 32 wide sdB binary systems have been obtained. The data are provided by high-resolution spectra gathered over more than a decade and a half.

Excluding the two outlier systems, EC03563-3618 and PG2148+095, the orbital period distribution of the ground-based wide sdB sample shows good agreement with the Galactic population synthesis model of \citet{Vos2020A&A}. These differences become more pronounced when the Gaia-wide sdB sample is included. 

Moreover, the Gaia sample shows an anti-correlation between orbital period and eccentricity. However, two factors prevent us from drawing further conclusions: the large uncertainties in the Gaia eccentricity measurements and the narrow orbital period range (700-900 days) where most systems are located. Whether this trend arises from an observational bias or represents a new population of wide sdB systems detected by Gaia remains a matter for further investigation.

Notably, no cross-matching records were obtained from the NSS files for sources compatible with a double-lined spectroscopic binary (SB2) model, even though the NSS files contain more than 5,000 records. This may be significant for the orbital parameters and features of the Gaia sample.
Several of the Gaia systems are currently being monitored. Gaia parameters and those derived from comparisons with high-resolution spectroscopy will help clarify this issue.

In contrast, the ground-based sample of wide sdB systems, excluding the two aforementioned outliers, does not show a clear linear relationship between orbital period and eccentricity. This contrasts with earlier studies that suggested otherwise and were based on fewer systems. 

In a forthcoming study \citep{Molina2026paperII}, the mass distributions of the companions in these systems will be derived using a method that combines spectroscopic, photometric, and evolutionary model approaches. The availability of absolute masses will provide other valuable parameters, including absolute separations and orbital inclinations. It will allow one to analyse the rotation periods of the companions and to identify entirely new parameter correlations, providing further strong constraints on binary evolution. 

We hope that this comprehensive set of wide sdB binaries will contribute to a better understanding of their formation and their main evolutionary characteristics.

\begin{acknowledgements}
J.V. acknowledges support from the Grant Agency of the Czech Republic (GA\v{C}R 22-34467S). MV  acknowledges funding support from the Fondecyt Regular project 1211491. A.B.~acknowledges support from the Australian Research Council (ARC) Centre of Excellence for Gravitational Wave Discovery (OzGrav), through project number CE230100016. This work has made use of data from the European Space Agency (ESA) mission {\it Gaia} (\url{https://www.cosmos.esa.int/gaia}), processed by the {\it Gaia} Data Processing and Analysis Consortium (DPAC, \url{https://www.cosmos.esa.int/web/gaia/dpac/consortium}). 
Funding for the DPAC has been provided by national institutions, in particular the institutions participating in the {\it Gaia} Multilateral Agreement.
Based on observations collected at the European Southern Observatory under ESO programs 088.D-0364, 093.D-0629, 096.D-0180, 097.D-0110, 098.D-0018, 099.D-0014, 0100.D-0082, 0101.D-0200, 0102.D-0255, 0103.D-0129, 0104.D-0135, 105.20L2.001 and 106.2105.001.
\end{acknowledgements}

\bibliographystyle{aa}
\bibliography{bibliography}


\begin{appendix} 

\onecolumn
\section{Observations, table}

\begin{table}
\caption{Summary of the main features of wide hot subdwarf binaries from both, Ground- and Gaia-based sample: ICRS coordinates (degrees, J2000), distances (from parallaxes), used spectrograph/TELESCOPE, number of observations, and photometric band magnitudes and RUWE index provided by Gaia DR3. Observations of Hermes, obtained from specific setups for blue or red interval spectra in different sessions, indicated as b/r, respectively. For the Gaia-based sample, their ground observations counterparts are also indicated.}
\label{21-Obs}
\begin{sideways}
\centering
\small
\renewcommand{\arraystretch}{1.1}
\begin{tabular}{|l@{\hskip 0.10in}l@{\hskip 0.10in}|r@{\hskip 0.10in}r@{\hskip 0.15in}|c@{\hskip 0.10in}|c@{\hskip 0.15in}c@{\hskip 0.15in}|c@{\hskip 0.15in}c@{\hskip 0.15in}c@{\hskip 0.15in}|c|}
\hline
\hline
Systems & Other id &  ICRS \,\,\, ra  & dec & dist. (pc) & Spect/TEL & Obs (n) & v (mag) &   g (mag) &  bp-rp (mag) & RUWE  \\
\hline
\multicolumn{11}{|l|}{ \textbf{Ground-based sample}}    \\
\hline
\footnotesize PG 0048+091               & \tiny                          & \footnotesize  12.86220  & \footnotesize   9.35914 & 1254.7   & \tiny uves/VLT           &  19     & 14.26	   &   14.18	  &    0.25	   &   1.01    \\ 
\footnotesize BD-11 162                 & \tiny                          & \footnotesize  13.06264  & \footnotesize -10.66294 & 331.2    & \tiny hermes/MERCATOR    &  56     & 11.18	   &   11.08	  &    0.01	   &   3.69    \\ 
\footnotesize PB 6355                   & \tiny                          & \footnotesize  19.11376  & \footnotesize   6.05317 & 766.3    & \tiny uves/VLT           &  19     & 13.04	   &   12.92	  &    0.41	   &   1.08    \\ 
\footnotesize MCT 0146-2651             & \tiny SB 744                   & \footnotesize  27.18394  & \footnotesize -26.60376 & 459.6    & \tiny uves/VLT           &  12     & 12.31	   &   12.22	  &    0.03	   &   2.60     \\ 
\footnotesize FAUST 321                 & \tiny J01513-7548              & \footnotesize  27.84755  & \footnotesize -75.81085 & 866.5    & \tiny uves/VLT           &  30     & 13.03	   &   12.95	  &    0.37	   &   1.11    \\ 
\footnotesize JL 277                    & \tiny                          & \footnotesize  30.39326  & \footnotesize -53.72875 & 869.0    & \tiny uves/VLT           &  30     & 13.09	   &   13.01	  &    0.23	   &   1.44     \\ 
\footnotesize GALEX J022836.7-362543    & \tiny UCAC4 268-002536         & \footnotesize  37.15355  & \footnotesize -36.42932 & 845.5    & \tiny uves/VLT           &  30     & 13.54	   &   13.43	  &    0.27	   &   1.67     \\ 
\footnotesize EC 03143-5945             & \tiny J03154-5934              & \footnotesize  48.87540  & \footnotesize -59.56803 & 986.4    & \tiny uves/VLT           &  26     & 13.38	   &   13.31	  &    0.21	   &   1.82    \\ 
\footnotesize GALEX J033216.7-023302    & \tiny J03322-0233              & \footnotesize  53.06977  & \footnotesize  -2.55056 & 981.4    & \tiny uves/VLT           &  26     & n.a.	   &   12.56	  &    0.41	   &   1.29    \\ 
\footnotesize EC 03563-3618             & \tiny J03582-3609              & \footnotesize  59.56180  & \footnotesize -36.16069 & 1513.6   & \tiny uves/VLT           &  24     & 13.60	   &   13.65	  &    0.45	   &   1.29    \\ 
\footnotesize HE 0430-2457              & \tiny                          & \footnotesize  68.26602  & \footnotesize -24.85548 & 978.4    & \tiny uves/VLT           &  29     & 14.12	   &   14.15	  &    0.05	   &   4.52    \\ 
\footnotesize GALEX J053939.1-283329    & \tiny UCAC2 19821041           & \footnotesize  84.91349  & \footnotesize -28.55862 & 1083.1   & \tiny uves/VLT           &  26     & 13.89	   &   13.80	  &    0.02	   &   2.15    \\ 
\footnotesize BD+34 1543                & \tiny                          & \footnotesize 107.53229  & \footnotesize  34.41467 & 194.4    & \tiny hermes/MERCATOR    &  30     & 10.17	   &   10.05	  &    0.33	   &   2.05    \\ 
\footnotesize GALEX J081110.8+273420    & \tiny                          & \footnotesize 122.79523  & \footnotesize  27.57236 & 701.3    & \tiny uves/VLT           &  17     & 12.42	   &   12.48	  &    0.23	   &   1.15    \\ 
\footnotesize J090001.55+012851.84      & \tiny SDSS J090001.54+012850.5 & \footnotesize 135.00646  & \footnotesize   1.48067 & 972.8    & \tiny uves/VLT           &  23     & n.a.	   &   13.07	  &    0.33	   &   1.00    \\ 
\footnotesize GALEX J101703.6-335502    & \tiny J10170-3355              & \footnotesize 154.26530  & \footnotesize -33.91725 & 723.6    & \tiny uves/VLT           &  21     & 12.91	   &   12.82	  &    0.34	   &   1.37    \\ 
\footnotesize PG 1018-047               & \tiny                          & \footnotesize 155.29403  & \footnotesize  -4.93880 & 808.5    & \tiny uves/VLT           &  21     & 13.31      &   13.25      &   -0.21	   &  3.48    \\ 
\footnotesize EC 11031-1348             & \tiny                          & \footnotesize 166.42254  & \footnotesize -14.07340 & 474.5    & \tiny hermes/MERCATOR    &  29/15  & 11.52	   &   11.43      &    0.44	   &   0.98    \\ 
\footnotesize PG 1104+243               & \tiny                          & \footnotesize 166.85899  & \footnotesize  24.05297 & 293.2    & \tiny hermes/MERCATOR    &  38     & 11.39      &   11.20      &    0.22	   &   2.30   \\ 
\footnotesize Feige 80                  & \tiny                          & \footnotesize 199.97331  & \footnotesize  12.06617 & 564.8    & \tiny hermes/MERCATOR    &  67     & 11.41	   &   11.26	  &   -0.04	   &   1.85    \\ 
\footnotesize Feige 87                  & \tiny                          & \footnotesize 205.06136  & \footnotesize  60.87959 & 446.5    & \tiny hermes/MERCATOR    &  33     & 11.71	   &   11.61	  &   -0.05	   &   3.73    \\ 
\footnotesize TYC 3871-835-1            & \tiny Balloon 82800003         & \footnotesize 228.90951  & \footnotesize  56.89549 & 462.1    & \tiny hermes/MERCATOR    &  56/18  & 11.42	   &   11.37	  &    0.32	   &   1.95    \\ 
\footnotesize PG 1514+034               & \tiny EGGR 440                 & \footnotesize 229.30947  & \footnotesize   3.17441 & 1187.1   & \tiny uves/VLT           &  22     & 13.99	   &   13.91	  &    0.12	   &   1.34   \\ 
\footnotesize BPS CS 22890-74           & \tiny BPS BS 16559-0077        & \footnotesize 231.01266  & \footnotesize   1.57266 & 1023.2   & \tiny uves/VLT           &  15     & 13.96	   &   13.87	  &    0.16	   &   1.56   \\ 
\footnotesize GALEX J162842.0+111838    & \tiny J16287+1118              & \footnotesize 247.17493  & \footnotesize  11.31105 & 921.1    & \tiny uves/VLT           &  12     & 13.22	   &   13.09	  &    0.42	   &   1.13   \\ 
\footnotesize TYC 2084-448-1            & \tiny GALEX J173651.2+280635   & \footnotesize 264.21339  & \footnotesize  28.10957 & 392.3    & \tiny hermes/MERCATOR    &  30/20  & 11.40	   &   11.45	  &    0.40	   &   1.31   \\ 
\footnotesize BD+29 3070                & \tiny BD+29 3070               & \footnotesize 264.58830  & \footnotesize  29.14658 & 293.0    & \tiny hermes/MERCATOR    &  31     & 10.45	   &   10.29	  &    0.34	   &   2.43   \\ 
\footnotesize GALEX J202216.8+015225    & \tiny                          & \footnotesize 305.56990  & \footnotesize   1.87341 & 720.7    & \tiny uves/VLT           &  18     & 13.43	   &   13.02	  &    0.43	   &   0.98    \\ 
\footnotesize BPS CS 22937-0067         & \tiny BPS CS 228937-67         & \footnotesize 318.07302  & \footnotesize -38.34854 & 1563.2   & \tiny uves/VLT           &  15     & 13.91	   &   13.80	  &    0.25	   &   1.00    \\ 
\footnotesize PG 2148+095               & \tiny                          & \footnotesize 327.82034  & \footnotesize   9.78318 & 1368.2   & \tiny uves/VLT           &  19     & 13.04	   &   12.94	  &    0.07	   &   1.27    \\ 
\footnotesize EC 22540-3324             & \tiny                          & \footnotesize 344.20349  & \footnotesize -33.14053 & 1520.4   & \tiny uves/VLT           &  15     & 13.86	   &   13.74	  &    0.45	   &   0.96   \\ 
\footnotesize BD -07 5977               & \tiny                          & \footnotesize 349.44497  & \footnotesize  -6.47523 & 775.3    & \tiny hermes/MERCATOR    &  45/18  & 10.54	   &   10.30	  &    0.88	   &   1.41    \\ 
\hline
\multicolumn{11}{|l|}{ \textbf{Gaia-based sample}}      \\
\hline
\footnotesize 392046852459641472     & \tiny UCAC4 683-002060          & \footnotesize   4.83582  & \footnotesize  46.55188  & 1567  & \tiny n.a.       &  n.a.     & n.a       &   14.25	&     0.24	   &   1.49    \\ 
\footnotesize 4850445797329363328    & \tiny CD-44 1028                & \footnotesize  48.31331  & \footnotesize -43.91643  &  291  & \tiny n.a.       &  n.a.     & 10.72     &   10.72	&     0.22	   &   2.08    \\ 
\footnotesize 972725503164737152     & \tiny UCAC4 718-043065          & \footnotesize  90.40099  & \footnotesize  53.48223  &  692  & \tiny n.a.       &  n.a.     & 14.09     &   13.93	&     0.22	   &   2.93    \\ 
\footnotesize 5579436712515286016    & \tiny UCAC4 275-013588          & \footnotesize 104.34083  & \footnotesize -35.19364  &  441  & \tiny uves/VLT   &  2        & 13.25     &   13.15	&     0.13	   &   5.55     \\ 
\footnotesize 3540092300847749760    & \tiny EC 11383-2238             & \footnotesize 175.21270  & \footnotesize -22.91281  &  834  & \tiny uves/VLT   &  7        & 13.31     &   13.15	&     0.27	   &   1.62    \\ 
\footnotesize 6335746093599431296    & \tiny PG 1459-048               & \footnotesize 225.52675  & \footnotesize  -4.99398  &  830  & \tiny uves/VLT   &  9        & 13.47     &   13.35	&     0.31	   &   1.48     \\ 
\footnotesize 4550114402362108416    & \tiny GALEX J172748.4+164456    & \footnotesize 261.95156  & \footnotesize  16.74908  &  839  & \tiny n.a.       &  n.a.     & 13.54     &   13.48	&     -0.05	   &   2.85     \\ 
\footnotesize 6359368722966483840    & \tiny                           & \footnotesize 270.50901  & \footnotesize -83.28525  &  992  & \tiny n.a.       &  n.a.     & n.a.      &   14.06	&     0.13	   &   1.96    \\ 
\footnotesize 4522995326025050496    & \tiny GALEX J182020.1+170933    & \footnotesize 275.08390  & \footnotesize  17.15881  & 1233  & \tiny n.a.       &  n.a.     & n.a.      &   14.67	&     0.12	   &   1.52    \\ 
\footnotesize 6631822855308840320    & \tiny                           & \footnotesize 283.26312  & \footnotesize -61.43595  & 1071  & \tiny n.a.       &  n.a.     & n.a.      &   14.57	&     0.29	   &   1.45    \\ 
\footnotesize 1806581377789928448    & \tiny GALEX J200739.6+134043    & \footnotesize 301.91495  & \footnotesize  13.67885  & 1090  & \tiny n.a.       &  n.a.     & n.a.      &   14.11	&     0.13	   &   2.03    \\ 
\footnotesize 2183496162809214464    & \tiny                           & \footnotesize 307.72960  & \footnotesize  54.38706  & 1500  & \tiny n.a.       &  n.a.     & n.a.      &   15.13	&     0.27	   &   2.16    \\ 
\footnotesize 6587335519633433472    & \tiny GALEX J220548.0-351939    & \footnotesize 331.45085  & \footnotesize -35.32744  &  786  & \tiny uves/VLT   &  9        & 13.25     &   13.18	&     0.01	   &   2.42    \\ 
\hline
\end{tabular}
\label{tb:21-Obs}
\end{sideways}
\end{table}

\onecolumn
\section{Orbital and Atmospheric Parameters, tables}

\begin{table}[h!]
\caption{Intervals and/or lines used in the CC with respect to the MS and sdB spectral mask models.} 
\label{tb:31-CCintervals}
\centering
\small
\renewcommand{\arraystretch}{1.24}
\begin{tabular}{|l@{\hskip 0.05in}|l@{\hskip 0.05in}|l|}
\hline
\hline
Systems                                     & CC MS intervals (\AA)                                                                                            &  CC sdB intervals / lines (\AA)                                \\
\Xhline{0.8pt}
\small \multirow{2}{*}{PG 0048+091}         &\tiny [3750, 4160] [4160, 4318] [4380, 4620] [4620, 4885]                                                                    &\tiny  \multirow{2}{*}{He I 5875}                                                    \\ 
                                            &\tiny  [4885, 4980] [5940, 6270] [6330, 6506]                                                                                &                                                                                     \\
\hline
\small \multirow{2}{*}{BD-11 162}           &\tiny [4725, 4820] [4895, 5385] [5440, 5870] [5880, 6270]                                                                    &\tiny  \multirow{2}{*}{He II 4686}                                                   \\ 
                                            &\tiny [6310, 6540]                                                                                                           &                                                                                     \\
\hline
\small \multirow{2}{*}{PB 6355}             &\tiny [4160, 4318] [4380, 4620] [4885, 4968] [5940, 6270]                                                                    &\tiny  [3920, 3930] [3980, 3990] [4178, 4188] [4250, 4286]                           \\ 
                                            &\tiny [6320, 6530]                                                                                                           & \tiny [4350, 4365] [4797, 4814]                                                      \\
\hline
\small \multirow{2}{*}{MCT 0146-2651}              &\tiny [3750, 4160] [4160, 4318] [4380, 4620] [4620, 4885]                                                                    &\tiny  \multirow{2}{*}{He I 5875}                                                     \\ 
                                            &\tiny [4885, 4980] [5700, 5940] [5940, 6270] [6320, 6530]                                                                    &                                                                                     \\
\hline
\small \multirow{2}{*}{Faust 321}           &\tiny [4150, 4313] [4520, 4620] [4690, 4815] [6055, 6205]                                                                    &\tiny  \multirow{2}{*}{He I 5875}                                                    \\ 
                                            &\tiny [6380, 6530]                                                                                                           &                                                                                     \\
\hline
\small \multirow{2}{*}{JL 277}              &\tiny [4180, 4300] [4400, 4430] [4485, 4675] [4725, 4825]                                                                    &\tiny  \multirow{2}{*}{[4030, 4080] [4525, 4575] [4625, 4654] [4708, 4718]}                            \\ 
                                            &\tiny [4935, 4980] [5910, 6270] [6330, 6510]                                                                                 &                                                                                     \\
\hline
\small GALEX J022836.7-362543               &\tiny [4885, 4980] [6096, 6180] [6388, 6466]                                                                                 &\tiny  [3990, 4000] [4413, 4419] [4628, 4632]                                          \\ 
\hline
\small \multirow{2}{*}{EC 03143-5945}       &\tiny [4180, 4300] [4485, 4675] [4725, 4825] [5885, 5960]                                                                    &\tiny  [4035, 4090] [4150, 4250] [4410, 4460] [4580, 4620]                              \\ 
                                            &\tiny [6050, 6200] [6380, 6470]                                                                                              & \tiny [4620, 4680]                                                                     \\
\hline
\small \multirow{2}{*}{GALEX J033216.7-023302} &\tiny [4117, 4318] [4160, 4318] [4380, 4620] [4885, 4968]                                                                 &\tiny  \multirow{2}{*}{[3993, 4048] [4235, 4288] [4580, 4620] [4620, 4680]}           \\ 
                                            &\tiny [5940, 6270] [6320, 6480]                                                                                              &                                                                                     \\
\hline
\small EC 03563-3618                        &\tiny [4400, 4468] [4475, 4675] [6000, 6260] [6330, 6450]                                                                   &\tiny  Triplet Si III 4553 \&  4568    \& 4574                                        \\ 

\hline
\small HE 0430-2457                         &\tiny [6060, 6200] [6387, 6480]                                                                                              &\tiny  [3993, 4010] [4410, 4460] [4580, 4620] [4620, 4680]                             \\ 
\hline
\small GALEX J053939.1-283329               &\tiny [4160, 4318] [4380, 4620] [4885, 4968] [5940, 6270]                                                                    &\tiny  He I 5875                                                                       \\ 
\hline
\small BD+34 1543                           &\tiny  [4780, 6530]                                                                            &\tiny  He I 5875                                                                       \\ 
\hline
\small GALEX J081110.8+273420                &\tiny [4420, 4465] [4485, 4675] [6000, 6260]                                                                                &\tiny  {He I 4471 \& He I 5875}                                                          \\ 
\hline
\small \multirow{2}{*}{SDSS J090001.54+012850.5} &\tiny [4160, 4318] [4380, 4620] [4885, 4968] [5940, 6270]                                                               &\tiny  \multirow{2}{*}{He I 5875}                                                     \\ 
                                            &\tiny [6320, 6530]                                                                                                           &                                                                                     \\
\hline
\small GALEX J101703.6-335502               &\tiny [4160, 4318] [4380, 4620] [4885, 4968]                                                                                 &\tiny  [4035, 4090] [4220, 4260] [4580, 4620] [4620, 4680]                              \\ 
\hline
\small \multirow{2}{*}{PG 1018-047}         &\tiny \multirow{2}{*}{[6095, 6460]}                                                                       &\tiny  [3910, 3957 ] [3990, 4082] [4149, 4308] [4364, 4477]                              \\ 
                                            &\tiny                                                                                                                        &\tiny  [4584, 4721]                                                                       \\
\hline
\small EC 11031-1348                        &\tiny  [4780, 6530]                                                                           &\tiny  He I 5875                                                                        \\ 
\hline
\small PG 1104+243                          &\tiny  [4780, 6530]                                                                           &\tiny  He I 5875                                                                        \\ 
\hline
\small \multirow{2}{*}{Feige 80}            &\tiny [4725 4820] [4895, 5385] [5440, 5870] [5880, 6270]                                                                     &\tiny  \multirow{2}{*}{He II 4686 \& He I 5875}                                         \\ 
                                            &\tiny [6310, 6540]                                                                                                           &                                                                                     \\
\hline
\small Feige 87                             &\tiny  [4780, 6530]                                                                           &\tiny  He I 5875                                                                       \\ 
\hline
\small TYC 3871-835-1                     &\tiny  [4780, 6530]                                                                           &\tiny  He I 5875                                                                       \\ 
\hline
\small \multirow{2}{*}{PG 1514+034}         &\tiny [4160, 4318] [4380, 4620] [4885, 4968] [5940, 6270]                                                                      &\tiny  \multirow{2}{*}{[4548, 4720] \& He I 5875}                                      \\ 
                                            &\tiny [6320, 6530]                                                                                                           &                                                                                     \\
\hline
\small \multirow{2}{*}{BPS CS 22890-74}   &\tiny [3750, 4160] [4160, 4318] [4380, 4620] [4620, 4885]                                                                    &\tiny  \multirow{2}{*}{[3990, 4080] [4400, 4450] [4480, 4560] \& He I 5875}           \\ 
                                            &\tiny [5940, 6270] [6320, 6530]                                                                                              &                                                                                     \\
\hline
\small GALEX J162842.0+111838               &\tiny [4400, 4468] [4475, 4675] [6000, 6260] [6330, 6450]                                                                   &\tiny  He I 5875                                            \\ 
\hline
\small GALEX J173651.2+280635               &\tiny  [4780, 6530]                                                                           &\tiny  He I 5875                                                                        \\ 
\hline
\small BD+29 3070                           &\tiny  [4780, 6530]                                                                           &\tiny  He I 5875                                                                        \\ 
\hline
\small \multirow{2}{*}{GALEX J202216.8+015225} &\tiny [3750, 4160] [4160, 4318] [4380, 4620] [4620, 4885]                                                                 &\tiny  \multirow{2}{*}{[4560, 4720] \& He I 5875}                                      \\ 
                                            &\tiny [4885, 4980] [5700, 5940] [5940, 6270] [6320, 6530] [6580, 6800]                                                       &                                                                                     \\
\hline
\small BPS CS 22937-0067                    &\tiny [4400, 4675] [6000, 6260]                                                                                 &\tiny  He II 4686                                                                        \\ 
\hline
\small \multirow{2}{*}{PG 2148+095}         &\tiny [4160, 4318] [4380, 4620] [4620, 4885] [4885, 4980]                                                                    &\tiny  \multirow{2}{*}{He I 5875}                                                                        \\ 
                                            &\tiny [5700, 5940] [5940, 6270] [6320, 6530]                                                                                 &                                                                                     \\
\hline
\small \multirow{2}{*}{EC 22540-3324}        &\tiny [4180, 4300] [4485, 4675] [4725, 4825] [5885, 5960]                                                                    &\tiny  [4035, 4090] [4183, 4188] [4410, 4460] [4580, 4620]                           \\ 
                                            &\tiny [6050, 6200] [6380, 6530]                                                                                              &\tiny  [4625, 4634]                                                                       \\
\hline
\small BD-07 5977                           &\tiny  [4780, 6530]                                                                           &\tiny  He I 5875                                                                          \\                  
\hline
\end{tabular}
\label{tb:31-CCintervals}

\end{table}

\begin{table}
\caption{Orbital parameters from the Keplerian orbit solutions of the wide hot subdwarf binary systems.} 
\label{tb:31-orbParam}
\begin{sideways}
\centering
\tiny
\renewcommand{\arraystretch}{1.6}
\begin{tabular}{|l@{\hskip 0.15in}|rl@{\hskip 0.1in}|rl@{\hskip 0.1in}|rl@{\hskip 0.15in}|rl@{\hskip 0.15in}|rl@{\hskip 0.15in}rl@{\hskip 0.1in}|rl@{\hskip 0.1in}rl@{\hskip 0.1in}|rl|}
\hline
\hline
Systems & \multicolumn{2}{l}{P(d) \hspace{0.2cm} $\sigma_{\text{P}}$(d)} &  \multicolumn{2}{|l}{T$_0$\,(d) \hspace{0.2cm} $\sigma_{\text{T0}}$(d)} & \multicolumn{2}{|l}{e \hspace{1.0cm} $\sigma_{\text{e}}$}  & \multicolumn{2}{|l}{$\omega$ \hspace{0.8cm} $\sigma_\omega$} & \multicolumn{2}{|l}{K$_{\rm MS}$\, (km s$^{-1}$)} &  \multicolumn{2}{l}{K$_{\rm sdB}$\, (km s$^{-1}$)} & \multicolumn{2}{|l}{$\gamma_{\rm MS}$\, (km s$^{-1}$)} &  \multicolumn{2}{l}{$\gamma_{\rm sdB}$\, (km s$^{-1}$)} &  \multicolumn{2}{|l|}{q \hspace{0.9cm} $\sigma_{\text{q}}$} \\
\hline
\footnotesize PG 0048+091               & 538  & \tiny $\pm 1$ & 2457489 & \tiny $\pm 7  $  & 0.072 & \tiny $\pm 0.008  $  & 0.44  &  \tiny $\pm 0.08  $ & 6.86  & \tiny $\pm 0.06 $  &  12.73  & \tiny $\pm  0.72  $   & -19.81   & \tiny $\pm 0.03 $ &-19.28  & \tiny $\pm 0.36 $ & 0.539 & \tiny $\pm 0.031 $       \\ 
\footnotesize BD-11 162                 & 497  & \tiny $\pm 1$ & 2454837 & \tiny $\pm 3  $  & 0.082 & \tiny $\pm 0.002  $  & 0.81  &  \tiny $\pm 0.01  $ & 8.71  & \tiny $\pm 0.02 $  &  18.00  & \tiny $\pm  0.50  $   & 3.29     & \tiny $\pm 0.01 $ &4.70    & \tiny $\pm 0.30 $ & 0.484 & \tiny $\pm 0.015 $       \\ 
\footnotesize PB 6355                   & 790  & \tiny $\pm 1$ & 2458092 & \tiny $\pm 15 $  & 0.127 & \tiny $\pm 0.007  $  & 3.34  &  \tiny $\pm 0.11  $ & 5.98  & \tiny $\pm 0.08 $  &  16.63  & \tiny $\pm  0.20  $   & -0.66    & \tiny $\pm 0.09 $ &3.24    & \tiny $\pm 0.24 $ & 0.360 & \tiny $\pm 0.007 $       \\ 
\footnotesize MCT 0146-2651             & 768  & \tiny $\pm 1$ & 2456080 & \tiny $\pm 4  $  & 0.094 & \tiny $\pm 0.023  $  & 5.39  &  \tiny $\pm 0.04  $ & 6.69  & \tiny $\pm 0.02 $  &  10.36  & \tiny $\pm  0.14  $   & 37.41    & \tiny $\pm 0.03 $ &41.78   & \tiny $\pm 0.14 $ & 0.646 & \tiny $\pm 0.009 $       \\ 
\footnotesize FAUST 321                 & 1014 & \tiny $\pm 1$ & 2455461 & \tiny $\pm 4  $  & 0.120 & \tiny $\pm 0.003  $  & 0.83  &  \tiny $\pm 0.03  $ & 5.84  & \tiny $\pm 0.01 $  &  13.08  & \tiny $\pm  0.15  $   & -38.43   & \tiny $\pm 0.01 $ &-37.23  & \tiny $\pm 0.13 $ & 0.447 & \tiny $\pm 0.005 $       \\ 
\footnotesize JL 277                    & 1090 & \tiny $\pm 1$ & 2455885 & \tiny $\pm 5  $  & 0.131 & \tiny $\pm 0.006  $  & 0.72  &  \tiny $\pm 0.03  $ & 6.46  & \tiny $\pm 0.04 $  &  15.72  & \tiny $\pm  0.07  $   & 102.81   & \tiny $\pm 0.04 $ &104.50  & \tiny $\pm 0.05 $ & 0.411 & \tiny $\pm 0.003 $       \\ 
\footnotesize GALEX J022836.7-362543    & 557  & \tiny $\pm 1$ & 2457128 & \tiny $\pm 2  $  & 0.174 & \tiny $\pm 0.004  $  & 5.00  &  \tiny $\pm 0.02  $ & 9.42  & \tiny $\pm 0.20 $  &  17.82  & \tiny $\pm  0.08  $   & -2.13    & \tiny $\pm 0.12 $ &-1.91   & \tiny $\pm 0.05 $ & 0.529 & \tiny $\pm 0.011 $       \\ 
\footnotesize EC 03143-5945             & 1034 & \tiny $\pm 2$ & 2457098 & \tiny $\pm 8  $  & 0.110 & \tiny $\pm 0.005  $  & 4.84  &  \tiny $\pm 0.04  $ & 7.27  & \tiny $\pm 0.04 $  &  16.97  & \tiny $\pm  0.07  $   & 39.91    & \tiny $\pm 0.03 $ &41.55   & \tiny $\pm 0.07 $ & 0.428 & \tiny $\pm 0.003 $       \\ 
\footnotesize GALEX J033216.7-023302    & 1257 & \tiny $\pm 2$ & 2457336 & \tiny $\pm 6  $  & 0.166 & \tiny $\pm 0.006  $  & 3.87  &  \tiny $\pm 0.04  $ & 6.46  & \tiny $\pm 0.04 $  &  18.62  & \tiny $\pm  0.11  $   & 27.57    & \tiny $\pm 0.03 $ &29.92   & \tiny $\pm 0.07 $ & 0.347 & \tiny $\pm 0.009 $       \\ 
\footnotesize EC 03563-3618             & 2092 & \tiny $\pm 4$ & 2457175 & \tiny $\pm 4  $  & 0.514 & \tiny $\pm 0.004  $  & 4.84  &  \tiny $\pm 0.01  $ & 4.52  & \tiny $\pm 0.04 $  &  14.03  & \tiny $\pm  0.11  $   & 10.00    & \tiny $\pm 0.02 $ &11.99   & \tiny $\pm 0.05 $ & 0.322 & \tiny $\pm 0.004 $       \\ 
\footnotesize HE 0430-2457              & 771  & \tiny $\pm 1$ & 2456267 & \tiny $\pm 47 $  & 0.015 & \tiny $\pm 0.005  $  & 4.08  &  \tiny $\pm 0.38  $ & 5.44  & \tiny $\pm 0.11 $  &  18.38  & \tiny $\pm  0.10  $   & 4.62     & \tiny $\pm 0.09 $ &6.25    & \tiny $\pm 0.09 $ & 0.296 & \tiny $\pm 0.006 $       \\ 
\footnotesize GALEX J053939.1-283329    & 842  & \tiny $\pm 1$ & 2455647 & \tiny $\pm 17 $  & 0.033 & \tiny $\pm 0.005  $  & 2.05  &  \tiny $\pm 0.12  $ & 6.72  & \tiny $\pm 0.03 $  &  9.80   & \tiny $\pm  0.24  $   & 12.85    & \tiny $\pm 0.03 $ &14.29   & \tiny $\pm 0.18 $ & 0.686 & \tiny $\pm 0.017 $       \\ 
\footnotesize BD+34 1543                & 972  & \tiny $\pm 2$ & 2451519 & \tiny $\pm 11 $  & 0.160 & \tiny $\pm 0.010  $  & 1.58  &  \tiny $\pm 0.07  $ & 5.91  & \tiny $\pm 0.07 $  &  10.31  & \tiny $\pm  0.22  $   & 32.10    & \tiny $\pm 0.06 $ &33.12   & \tiny $\pm 0.15 $ & 0.574 & \tiny $\pm 0.011 $       \\ 
\footnotesize GALEX J081110.8+273420    & 773  & \tiny $\pm 1$ & 2457409 & \tiny $\pm 6  $  & 0.136 & \tiny $\pm 0.011  $  & 4.09  &  \tiny $\pm 0.05  $ & 3.49  & \tiny $\pm 0.03 $  &  6.97   & \tiny $\pm  0.21  $   & 45.29    & \tiny $\pm 0.02 $ &45.08   & \tiny $\pm 0.16 $ & 0.501 & \tiny $\pm 0.016 $       \\ 
\footnotesize J090001.55+012851.84      & 1132 & \tiny $\pm 3$ & 2456847 & \tiny $\pm 1  $  & 0.192 & \tiny $\pm 0.011  $  & 0.17  &  \tiny $\pm 0.03  $ & 6.79  & \tiny $\pm 0.16 $  &  16.06  & \tiny $\pm  0.22  $   & 24.80    & \tiny $\pm 0.09 $ &25.65   & \tiny $\pm 0.11 $ & 0.422 & \tiny $\pm 0.012 $       \\ 
\footnotesize GALEX J101703.6-335502    & 1073 & \tiny $\pm 2$ & 2457769 & \tiny $\pm 14 $  & 0.112 & \tiny $\pm 0.010  $  & 3.57  &  \tiny $\pm 0.09  $ & 1.92  & \tiny $\pm 0.03 $  &  5.19   & \tiny $\pm  0.09  $   & -2.80    & \tiny $\pm 0.01 $ &-1,07   & \tiny $\pm 0.05 $ & 0.370 & \tiny $\pm 0.008 $       \\ 
\footnotesize PG 1018-047               & 752  & \tiny $\pm 2$ & 2455193 & \tiny $\pm 16 $  & 0.050 & \tiny $\pm 0.010  $  & 1.60  &  \tiny $\pm 0.20  $ & 6.95  & \tiny $\pm 0.40 $  &  10.46  & \tiny $\pm  0.09  $   & 38.38    & \tiny $\pm 0.06 $ &37.82   & \tiny $\pm 0.19 $ & 0.664 & \tiny $\pm 0.039 $       \\ 
\footnotesize EC 11031-1348             & 1099 & \tiny $\pm 6$ & 2456600 & \tiny $\pm 20 $  & 0.170 & \tiny $\pm 0.030  $  & 3.90  &  \tiny $\pm 0.10  $ & 5.55  & \tiny $\pm 0.15 $  &  15.50  & \tiny $\pm  0.60  $   & -14.75   & \tiny $\pm 0.08 $ &-11.40  & \tiny $\pm 0.30 $ & 0.358 & \tiny $\pm 0.017 $       \\ 
\footnotesize PG 1104+243               & 755  & \tiny $\pm 3$ & 2450480 & \tiny $\pm 8  $  & 0.040 & \tiny $\pm 0.020  $  & 0.70  &  \tiny $\pm 0.02  $ & 4.42  & \tiny $\pm 0.04 $  &  6.34   & \tiny $\pm  0.20  $   & -15.59   & \tiny $\pm 0.06 $ &-13.80  & \tiny $\pm 0.20 $ & 0.697 & \tiny $\pm 0.023 $       \\ 
\footnotesize Feige 80                  & 1140 & \tiny $\pm 5$ & 2454430 & \tiny $\pm 7  $  & 0.160 & \tiny $\pm 0.020  $  & 5.96  &  \tiny $\pm 0.02  $ & 6.20  & \tiny $\pm 0.20 $  &  15.00  & \tiny $\pm  0.70  $   & 40.21    & \tiny $\pm 0.07 $ &39.90   & \tiny $\pm 0.70 $ & 0.413 & \tiny $\pm 0.023 $       \\ 
\footnotesize Feige 87                  & 936  & \tiny $\pm 2$ & 2453259 & \tiny $\pm 21 $  & 0.110 & \tiny $\pm 0.010  $  & 2.92  &  \tiny $\pm 0.15  $ & 8.19  & \tiny $\pm 0.11 $  &  15.01  & \tiny $\pm  0.21  $   & 32.98    & \tiny $\pm 0.08 $ &34.32   & \tiny $\pm 0.16 $ & 0.546 & \tiny $\pm 0.011 $       \\ 
\footnotesize TYC 3871-835-1            & 1263 & \tiny $\pm 5$ & 2454075 & \tiny $\pm 18 $  & 0.160 & \tiny $\pm 0.020  $  & 2.83  &  \tiny $\pm 0.08  $ & 2.31  & \tiny $\pm 0.04 $  &  4.24   & \tiny $\pm  0.20  $   & -14.98   & \tiny $\pm 0.02 $ &-13.36  & \tiny $\pm 0.12 $ & 0.545 & \tiny $\pm 0.027 $       \\ 
\footnotesize PG 1514+034               & 480  & \tiny $\pm 1$ & 2456741 & \tiny $\pm 3  $  & 0.082 & \tiny $\pm 0.004  $  & 0.44  &  \tiny $\pm 0.04  $ & 10.18 & \tiny $\pm 0.03 $  &  17.72  & \tiny $\pm  0.20  $   & -71.50   & \tiny $\pm 0.03 $ &-70.93  & \tiny $\pm 0.14 $ & 0.574 & \tiny $\pm 0.007 $      \\ 
\footnotesize BPS CS 22890-74           & 1225 & \tiny $\pm 1$ & 2456583 & \tiny $\pm 5  $  & 0.144 & \tiny $\pm 0.005  $  & 0.01  &  \tiny $\pm 0.03  $ & 6.19  & \tiny $\pm 0.03 $  &  8.32   & \tiny $\pm  0.22  $   & -1.45    & \tiny $\pm 0.02 $ &0.76    & \tiny $\pm 0.12 $ & 0.744 & \tiny $\pm 0.020 $      \\ 
\footnotesize GALEX J162842.0+111838    & 871  & \tiny $\pm 1$ & 2457547 & \tiny $\pm 13 $  & 0.101 & \tiny $\pm 0.015  $  & 6.13  &  \tiny $\pm 0.11  $ & 4.21  & \tiny $\pm 0.07 $  &  7.78   & \tiny $\pm  0.43  $   & -42.69   & \tiny $\pm 0.03 $ &-44.18  & \tiny $\pm 0.22 $ & 0.540 & \tiny $\pm 0.031 $      \\ 
\footnotesize TYC 2084-448-1            & 1098 & \tiny $\pm 5$ & 2456054 & \tiny $\pm 58 $  & 0.050 & \tiny $\pm 0.030  $  & 5.63  &  \tiny $\pm 0.33  $ & 6.30  & \tiny $\pm 0.10 $  &  12.25  & \tiny $\pm  0.30  $   & -15.56   & \tiny $\pm 0.06 $ &-13.30  & \tiny $\pm 0.20 $ & 0.514 & \tiny $\pm 0.015 $      \\ 
\footnotesize BD+29 3070                & 1254 & \tiny $\pm 5$ & 2453877 & \tiny $\pm 41 $  & 0.150 & \tiny $\pm 0.010  $  & 1.60  &  \tiny $\pm 0.22  $ & 6.53  & \tiny $\pm 0.30 $  &  16.60  & \tiny $\pm  0.60  $   & -57.58   & \tiny $\pm 0.36 $ &-56.80  & \tiny $\pm 0.90 $ & 0.393 & \tiny $\pm 0.023 $      \\ 
\footnotesize GALEX J202216.8+015225    & 923  & \tiny $\pm 1$ & 2457619 & \tiny $\pm 5  $  & 0.198 & \tiny $\pm 0.005  $  & 2.67  &  \tiny $\pm 0.04  $ & 6.18  & \tiny $\pm 0.04 $  &  11.35  & \tiny $\pm  0.27  $   & -32.19   & \tiny $\pm 0.04 $ &-31.53  & \tiny $\pm 0.21 $ & 0.544 & \tiny $\pm 0.014 $       \\ 
\footnotesize BPS CS 22937-0067         & 1113 & \tiny $\pm 2$ & 2456604 & \tiny $\pm 23 $  & 0.120 & \tiny $\pm 0.018  $  & 3.68  &  \tiny $\pm 0.12  $ & 6.40  & \tiny $\pm 0.12 $  &  15.23  & \tiny $\pm  0.77  $   & -7.90    & \tiny $\pm 0.07 $ &-9.73   & \tiny $\pm 0.49 $ & 0.420 & \tiny $\pm 0.023 $       \\ 
\footnotesize PG 2148+095               & 1558 & \tiny $\pm 3$ & 2457342 & \tiny $\pm 1  $  & 0.361 & \tiny $\pm 0.009  $  & 5.36  &  \tiny $\pm 0.01  $ & 7.72  & \tiny $\pm 0.12 $  &  16.27  & \tiny $\pm  0.45  $   & -145.66  & \tiny $\pm 0.04 $ &-142.13 & \tiny $\pm 0.24 $ & 0.474 & \tiny $\pm 0.015 $       \\ 
\footnotesize EC 22540-3324             & 633  & \tiny $\pm 1$ & 2455993 & \tiny $\pm 21 $  & 0.047 & \tiny $\pm 0.006  $  & 3.53  &  \tiny $\pm 0.22  $ & 8.65  & \tiny $\pm 0.11 $  &  13.53  & \tiny $\pm  0.13  $   & 28.53    & \tiny $\pm 0.09 $ &28.98   & \tiny $\pm 0.11 $ & 0.639 & \tiny $\pm 0.010 $      \\ 
\footnotesize BD-07 5977                & 1262 & \tiny $\pm 1$ & 2454971 & \tiny $\pm 4  $  & 0.160 & \tiny $\pm 0.010  $  & 5.50  &  \tiny $\pm 0.10  $ & 2.62  & \tiny $\pm 0.01 $  &  6.20   & \tiny $\pm  1.00  $   & -8.62    & \tiny $\pm 0.01 $ &-5.50   & \tiny $\pm 0.03 $ & 0.423 & \tiny $\pm 0.068 $       \\

\hline
\end{tabular}
\label{tb:31-orbParam}
\end{sideways}
\end{table}

\begin{table}
\caption{Atmospheric parameters from spectroscopic analysis by \gssp of cool companions in the wide hot subdwarf binary systems from both Ground-based and Gaia-based observations when Ground observations are available. The Gaia systems shown are part of an ongoing observation program (116.28ZZ.001). Included 1$\sigma$ error intervals.} 
\label{GSSP}
\centering
\small
\renewcommand{\arraystretch}{1.6}
\begin{tabular}{l@{\hskip 0.5in}rl@{\hskip 0.2in}rl@{\hskip 0.4in}rl@{\hskip 0.3in}rl@{\hskip 0.3in}rl}
\hline
\hline
Systems & \multicolumn{2}{l}{\teff (K)} &  \multicolumn{2}{l}{\met (dex)} & \multicolumn{2}{l}{\logg (dex)} & \multicolumn{2}{l}{\vsin(\kms)} &  \multicolumn{2}{l}{Dilution}\\
\hline
\multicolumn{11}{l}{ \textbf{Ground-based sample}}    \\
\hline
\small PG 0048+091               & 6471   & \tiny $^{+367}_{-388}$   & -0.47  & \tiny $\pm 0.21 $          & 4.6   & \tiny $\pm 0.6  $  & 26 &  \tiny $\pm 2$ &  0.68  &  \tiny   $\pm 0.14 $    \\ 
\small BD-11 162                 & 5630   & \tiny $^{+230}_{-170}$   & -0.6   & \tiny $\pm 0.20 $          & 4.3   & \tiny $\pm 0.5  $  & 12 &  \tiny $\pm 1$ &  0.52  &  \tiny   $\pm 0.08 $    \\ 
\small PB 6355                   & 6516   & \tiny $\pm 146  $        & -0.19  & \tiny $\pm 0.09 $          & 4.5   & \tiny $\pm 0.1  $  & 57 &  \tiny $\pm 2$ &  0.85  &  \tiny   $\pm 0.07 $    \\ 
\small MCT 0146-2651             & 5792   & \tiny $^{+305}_{-285}$   & -1.09  & \tiny $^{+0.20}_{-0.24}$   & 4.6   & \tiny $\pm 0.5  $  & 5  &  \tiny $\pm 2$ &  0.53  &  \tiny   $\pm 0.11 $     \\ 
\small FAUST 321                 & 6663   & \tiny $\pm 119  $        & -0.24  & \tiny $\pm 0.07 $          & 4.5   & \tiny $\pm 0.1  $  & 26 &  \tiny $\pm 1$ &  0.77  &  \tiny   $\pm 0.06 $    \\ 
\small JL 277                    & 6694   & \tiny $\pm 245  $        & -0.31  & \tiny $\pm 0.10 $          & 5.0   & \tiny $\pm 0.3  $  & 32 &  \tiny $\pm 2$ &  0.64  &  \tiny   $\pm 0.12 $     \\ 
\small GALEX J022836.7-362543    & 5330   & \tiny $\pm 180  $        & -0.15  & \tiny $\pm 0.15 $          & 4.5   & \tiny $\pm 0.2  $  & 90 &  \tiny $\pm 5$ &  0.30  &  \tiny   $\pm 0.05 $     \\ 
\small EC 03143-5945             & 6080   & \tiny $\pm 190  $        & -0.30  & \tiny $^{+0.15}_{-0.10}$   & 4.4   & \tiny $\pm 0.3  $  & 23 &  \tiny $\pm 1$ &  0.55  &  \tiny   $\pm 0.02 $    \\ 
\small GALEX J033216.7-023302    & 7108   & \tiny $\pm 184  $        & -0.16  & \tiny $\pm 0.10 $          & 4.2   & \tiny $\pm 0.4  $  & 99 &  \tiny $\pm 4$ &  1.00  &  \tiny   $\pm 0.03 $    \\ 
\small EC 03563-3618             & 6143   & \tiny $^{+229}_{-200}$   &  0.04  & \tiny $\pm 0.16 $          & 4.4   & \tiny $\pm 0.5  $  & 29 &  \tiny $\pm 2$ &  0.90  &  \tiny   $\pm 0.14 $    \\ 
\small HE 0430-2457              & 4620   & \tiny $^{+170}_{-200}$   & -0.35  & \tiny $\pm 0.30 $          & 4.5   & \tiny $\pm 0.5  $  & 30 &  \tiny $\pm 3$ &  0.20  &  \tiny   $\pm 0.05 $    \\ 
\small GALEX J053939.1-283329    & 5582   & \tiny $\pm 163  $        & -0.33  & \tiny $\pm 0.12 $          & 4.5   & \tiny $\pm 0.2  $  & 7  &  \tiny $\pm 1$ &  0.39  &  \tiny   $\pm 0.05 $    \\ 
\small BD+34 1543                & 6100   & \tiny $\pm 150  $        & -0.31  & \tiny $\pm 0.10 $          & 4.5   & \tiny $\pm 0.5  $  & 19 &  \tiny $\pm 1$ &  0.60  &  \tiny   $\pm 0.05 $    \\ 
\small GALEX J081110.8+273420    & 7031   & \tiny $^{+224}_{-286}$   & -0.63  & \tiny $\pm 0.12 $          & 5.0   & \tiny $\pm 0.3  $  & 10 &  \tiny $\pm 1$ &  0.72  &  \tiny   $\pm 0.12 $    \\ 
\small J090001.55+012851.84      & 6604   & \tiny $^{+292}_{-265}$   & -0.23  & \tiny $\pm 0.14 $          & 4.5   & \tiny $\pm 0.4  $  & 46 &  \tiny $\pm 3$ &  0.90  &  \tiny   $\pm 0.10 $    \\ 
\small GALEX J101703.6-335502    & 6306   & \tiny $\pm 135  $        & -0.19  & \tiny $\pm 0.08 $          & 4.5   & \tiny $\pm 0.2  $  & 19 &  \tiny $\pm 1$ &  0.69  &  \tiny   $\pm 0.04 $    \\ 
\small PG 1018-047               & 4461   & \tiny $\pm 24  $         & -0.54  & \tiny $\pm 0.16 $          & 4.1   & \tiny $\pm 0.1  $  & 19 &  \tiny $\pm 1$ &  0.10  &  \tiny   $\pm 0.03 $    \\ 
\small EC 11031-1348             & 6355   & \tiny $\pm 257  $        &  0.11  & \tiny $\pm 0.13 $          & 4.3   & \tiny $\pm 0.5  $  & 70 &  \tiny $\pm 4$ &  0.84  &  \tiny   $\pm 0.10 $    \\ 
\small PG 1104+243               & 5851   & \tiny $\pm 165  $        & -0.58  & \tiny $\pm 0.12 $          & 4.4   & \tiny $\pm 0.2  $  & 12 &  \tiny $\pm 1$ &  0.57  &  \tiny   $\pm 0.06 $    \\ 
\small Feige 80                  & 6200   & \tiny $^{+320}_{-290}$   & -0.50  & \tiny $\pm 0.20 $          & 3.9   & \tiny $\pm 0.6  $  & 29 &  \tiny $\pm 2$ &  0.55  &  \tiny   $\pm 0.11 $    \\ 
\small Feige 87                  & 5500   & \tiny $\pm 230  $        & -0.50  & \tiny $\pm 0.15 $          & 4.5   & \tiny $\pm 0.5  $  & 8  &  \tiny $\pm 1$ &  0.30  &  \tiny   $\pm 0.05 $    \\ 
\small TYC 3871-835-1            & 6216   & \tiny $\pm 196  $        &  0.09  & \tiny $\pm 0.10 $          & 4.3   & \tiny $\pm 0.3  $  & 15 &  \tiny $\pm 1$ &  0.61  &  \tiny   $\pm 0.04 $    \\ 
\small PG 1514+034               & 5704   & \tiny $^{+245}_{-214}$   & -0.58  & \tiny $^{+0.17}_{-0.20}$   & 4.5   & \tiny $\pm 0.4  $  & 18 &  \tiny $\pm 1$ &  0.54  &  \tiny   $\pm 0.12 $   \\ 
\small BPS CS 22890-74           & 5630   & \tiny $^{+170}_{-220}$   &  0.10  & \tiny $^{+0.20}_{-0.10}$   & 4.5   & \tiny $\pm 0.2  $  & 7  &  \tiny $\pm 1$ &  0.45  &  \tiny   $\pm 0.05 $   \\ 
\small GALEX J162842.0+111838    & 6916   & \tiny $^{+239}_{-292}$   & -0.37  & \tiny $^{+0.12}_{-0.14}$   & 4.9   & \tiny $\pm 0.4  $  & 21 &  \tiny $\pm 1$ &  0.84  &  \tiny   $\pm 0.14 $   \\ 
\small TYC 2084-448-1            & 6229   & \tiny $\pm 229  $        & -0.06  & \tiny $\pm 0.13 $          & 4.5   & \tiny $\pm 0.4  $  & 52 &  \tiny $\pm 3$ &  0.76  &  \tiny   $\pm 0.05 $   \\ 
\small BD+29 3070                & 6170   & \tiny $\pm 200  $        &  0.05  & \tiny $\pm 0.10 $          & 4.3   & \tiny $\pm 0.4  $  & 52 &  \tiny $\pm 2$ &  0.70  &  \tiny   $\pm 0.04 $   \\ 
\small GALEX J202216.8+015225    & 6449   & \tiny $^{+265}_{-286}$   & -0.38  & \tiny $^{+0.15}_{-0.16}$   & 4.7   & \tiny $\pm 0.4  $  & 34 &  \tiny $\pm 2$ &  0.80  &  \tiny   $\pm 0.14 $    \\ 
\small BPS CS 22937-0067         & 7008   & \tiny $^{+395}_{-465}$   & -0.55  & \tiny $^{+0.18}_{-0.32}$   & 4.9   & \tiny $\pm 0.6  $  & 37 &  \tiny $\pm 3$ &  0.74  &  \tiny   $\pm 0.20 $    \\ 
\small PG 2148+095               & 6537   & \tiny $\pm 208  $        & -0.47  & \tiny $\pm 0.13 $          & 4.5   & \tiny $\pm 0.1  $  & 24 &  \tiny $\pm 1$ &  0.45  &  \tiny   $\pm 0.07 $    \\ 
\small EC 22540-3324             & 5449   & \tiny $^{+286}_{-245}$   & -0.62  & \tiny $^{+0.24}_{-0.23}$   & 4.0   & \tiny $\pm 0.5  $  & 59 &  \tiny $\pm 3$ &  0.88  &  \tiny   $\pm 0.16 $   \\ 
\small BD-07 5977                & 4792   & \tiny $\pm 98 $          & -0.16  & \tiny $\pm 0.12 $          & 3.0   & \tiny $\pm 0.4  $  & 9  &  \tiny $\pm 1$ &  0.72  &   \tiny  $\pm 0.04 $    \\
\hline
\multicolumn{11}{l}{ \textbf{Available Gaia-based sample (Ground observations)}}    \\
\hline

\small EC 11383-2238                & 5669   & \tiny $\pm 204 $         & -0.03  & \tiny $^{+0.16}_{-0.18}$    & 4.3   & \tiny $\pm 0.5  $  & 10   &  \tiny $\pm 1$        &  0.68 &  \tiny     $\pm 0.10 $    \\ 
\small PG 1459-048                  & 6429   & \tiny $^{+204}_{-229}$   & -0.43  & \tiny $\pm 0.12 $           & 5.0   & \tiny $\pm 0.4  $  & 5    &  \tiny $^{+1}_{-2}$   &  0.62 &  \tiny     $\pm 0.08 $   \\ 
\small GALEX J220548.0-351939       & 5914   & \tiny $\pm 294 $         & -0.57  & \tiny $^{+0.18}_{-0.20}$    & 4.7   & \tiny $\pm 0.5  $  & 2    &  \tiny $\pm 2 $       &  0.43 &  \tiny     $\pm 0.06 $    \\

\hline

\end{tabular}
\label{tb:GSSP}
\end{table}

\end{appendix}

\end{document}